\documentclass[12pt,thmsa,notitlepage]{article}
\usepackage{sw20lart}

\input{tcilatex}
\begin{document}

\author{Ling-Fong Li$^{*}$ and T. P. Cheng$^{\dagger }$ \and $^{*}${\small %
Department of Physics, Carnegie Mellon University} \and {\small Pittsburgh,
PA 15213, USA} \and $^{\dagger }${\small Department of Physics and
Astronomy, University of Missouri} \and {\small St. Louis, MO 63121, USA}}
\title{The Proton Spin and Flavor Structure in the Chiral Quark Model}
\date{Schladming Lectures (March 1997)}
\maketitle

\begin{abstract}
After a pedagogical review of the simple constituent quark model and deep
inelastic sum rules, we describe how a quark sea as produced by the emission
of internal Goldstone bosons by the valence quarks can account for the
observed features of proton spin and flavor structures. Some issues
concerning the strange quark content of the nucleon are also discussed.
\end{abstract}

\tableofcontents

\newpage

We shall first recall the contrasting concepts of current quarks \emph{vs}
constituent quarks. In the first Section we also briefly review the
successes and inadequacies of the simple constituent quark model (sQM) which
attempts to describe the properties of light hadrons as a composite systems
of $u,d,$ and $s$ valence quarks. Some of the more prominent features,
gleaned from the mass and spin systematics, are discussed. In the Sec. 2 we
shall provide a pedagogical review of the deep inelastic sum rules that can
be derived by way of operator product expansion and/or the simple parton
model. We show in particular how some the sum rules in the second category
can be interpreted as giving information of the nucleon quark sea. In the
remainder of these lectures we shall show that the account of the quark sea
as given by the chiral quark model is in broad agreement with the
experimental observation.

\section{Strong Interaction Symmetries and the Quark Model}

In the approximation of neglecting the light quark masses, the QCD
Lagrangian has the global $SU\left( 3\right) _{L}\times SU\left( 3\right)
_{R}$ symmetry. Namely, it is invariant under independent $SU\left( 3\right) 
$ transformation of the three left-handed and right-handed light quark
fields. This symmetry is realized in the Nambu-Goldstone mode with the
ground state being symmetric only with respect to the vector $SU\left(
3\right) _{L+R}$ transformations. This gives rise to an octet of Goldstone
bosons, which are identified with the low lying pseudoscalar mesons $\left(
\pi ,\,K,\,\eta \right) .$ For a pedagogical review see, for example, ref.%
\cite{CLbook}.

\subsection{Current quark mass ratios as deduced from pseudoscalar meson
masses}

The light current quark masses are the chiral symmetry breaking parameters
of the QCD Lagrangian. Their relative magnitude can be deduced from the soft
meson theorems for the pseudoscalar meson masses.

The matrix element of an axial vector current operator $A_{\mu }^{a}\;$taken
between the vacuum and one meson $\phi ^{b}$ state$\;$(with momentum $k_{\mu
}$) defines the decay constant $f_{a}$ as 
\[
\left\langle 0\left| A_{\mu }^{a}\right| \phi ^{b}\left( k\right)
\right\rangle =ik_{\mu }f_{a}\delta _{ab} 
\]
where the SU(3) indices $a,b...$range from $1,2,....8.\;$This means that the
divergence of the axial vector current has matrix element of 
\begin{equation}
\left\langle 0\left| \partial ^{\mu }A_{\mu }^{a}\right| \phi ^{b}\left(
k\right) \right\rangle =m_{a}^{2}f_{a}\delta _{ab},
\end{equation}
If the axial divergences are good interpolating fields for the pseudoscalar
mesons, we have the result of PCAC: 
\begin{equation}
\partial ^{\mu }A_{\mu }^{a}=m_{a}^{2}f_{a}\phi ^{a}.
\end{equation}
Using PCAC and the reduction formula we can derive a soft-meson theorem for
the pseudoscalar meson masses: 
\begin{eqnarray}
m_{a}^{2}f_{a}^{2}\delta _{ab} &=&-i\int d^{4}xe^{-ik\cdot x}\left\langle
0\left| \delta \left( x_{0}\right) \left[ A_{0}^{b}\left( x\right) ,\partial
^{\mu }A_{\mu }^{a}\left( 0\right) \right] \right| 0\right\rangle  \nonumber
\\
&=&-\left\langle 0\left| \left[ Q^{5b},\left[ Q^{5a},\mathcal{H}\left(
0\right) \right] \right] \right| 0\right\rangle  \label{sigmass}
\end{eqnarray}
where the axial charge is related to the time component of the axial vector
current as $Q^{5a}=\int d^{3}xA_{0}^{a}\left( x\right) .$

If we neglect the electromagnetic radiative correction, only the quark
masses, 
\begin{equation}
\mathcal{H}_{m}=m_{u}\bar{u}u+m_{d}\bar{d}d+m_{s}\bar{s}s,  \label{Hm}
\end{equation}
break the chiral symmetry. Hence only such terms are relevant in the
computation of the above commutators. [In actual computation it is simpler
if one takes $\mathcal{H}_{m}$ and $Q^{5a}$ to be $3\times 3$ matrices and
compute directly the \emph{anticommutator }in $\left[ \bar{q}\frac{\lambda
^{a}}{2}\gamma _{0}\gamma _{5}q,\,\bar{q}\lambda ^{b}q\right] =-\frac{1}{2}%
\bar{q}\left\{ \lambda ^{a},\,\lambda ^{b}\right\} \gamma _{5}q$.] In this
way we obtain: 
\begin{eqnarray}
f_{\pi }^{2}m_{\pi }^{2} &=&\frac{1}{2}\left( m_{u}+m_{d}\right)
\left\langle 0\left| \left( \bar{u}u+\bar{d}d\right) \right| 0\right\rangle 
\nonumber \\
f_{K}^{2}m_{K}^{2} &=&\frac{1}{2}\left( m_{u}+m_{s}\right) \left\langle
0\left| \left( \bar{u}u+\bar{s}s\right) \right| 0\right\rangle
\label{psmass1} \\
f_{\eta }^{2}m_{\eta }^{2} &=&\frac{1}{6}\left( m_{u}+m_{d}\right)
\left\langle 0\left| \left( \bar{u}u+\bar{d}d\right) \right| 0\right\rangle +%
\frac{4}{3}m_{s}\left\langle 0\left| \bar{s}s\right| 0\right\rangle . 
\nonumber
\end{eqnarray}

\subsubsection{Gell-Mann-Okubo mass relation and the strange to non-strange
quark mass ratio}

Since the flavor SU(3) symmetry is not spontaneously broken, 
\begin{equation}
\left\langle 0\left| \bar{u}u\right| 0\right\rangle =\left\langle 0\left| 
\bar{d}d\right| 0\right\rangle =\left\langle 0\left| \bar{s}s\right|
0\right\rangle \equiv \mu ^{3}
\end{equation}
and $f_{\pi }=f_{K}=f_{\eta }\equiv f;$ Eq.(\ref{psmass1}) is simplified to 
\begin{eqnarray}
m_{\pi }^{2} &=&2m_{n}\frac{\mu ^{3}}{f^{2}}\;  \nonumber \\
m_{K}^{2} &=&\left( m_{n}+m_{s}\right) \frac{\mu ^{3}}{f^{2}}  \nonumber \\
m_{\eta }^{2} &=&\frac{2}{3}\left( m_{n}+2m_{s}\right) \frac{\mu ^{3}}{f^{2}}
\label{psmass}
\end{eqnarray}
where we have made the approximation of $m_{u}\simeq m_{d}\equiv m_{n}.$
From this, we can deduce the Gell-Mann-Okubo mass relation for the 0$^{-}$
mesons: 
\begin{equation}
3m_{\eta }^{2}=4m_{K}^{2}-m_{\pi }^{2},  \label{GMOeta}
\end{equation}
as well as the strange to nonstrange quark mass ratio \cite{GMOR}: 
\begin{equation}
\frac{m_{n}}{m_{s}}=\frac{m_{u}+m_{d}}{2m_{s}}=\frac{m_{\pi }^{2}}{%
2m_{K}^{2}-m_{\pi }^{2}}\simeq \frac{1}{25}.  \label{n-s-ratio}
\end{equation}

\subsubsection{Isospin breaking by the strong interaction \& the $%
m_{u}/m_{d} $ ratio}

In order to study the ratio of $m_{u}/m_{d}$, we need to include the
electromagnetic radiative contribution to the masses. The effective
Hamiltonian due to virtual photon exchange is given by 
\begin{equation}
\mathcal{H}_{\gamma }=e^{2}\int d^{4}xT\left( J_{\mu }^{em}\left( x\right)
J_{\nu }^{em}\left( 0\right) \right) D^{\mu \nu }\left( x\right) 
\label{Hgamma}
\end{equation}
where $D^{\mu \nu }\left( x\right) $ is the photon propagator. Thus, beside
the contribution from $\mathcal{H}_{m},$ we also have the additional term on
the RHS of eqn.(\ref{sigmass}): 
\begin{equation}
\sigma _{\gamma }^{ab}=\left\langle 0\left| \left[ Q^{5b},\left[ Q^{5a},%
\mathcal{H}_{\gamma }\right] \right] \right| 0\right\rangle 
\end{equation}
Now we make the observation (\emph{Dashen's theorem}\cite{Dashen})\emph{\ : }%
For the electrically neutral mesons, we have $\left[ Q^{5a},\mathcal{H}%
_{\gamma }\right] =0,$ which leads to 
\begin{equation}
\sigma _{\gamma }\left( \pi ^{0}\right) =\sigma _{\gamma }\left(
K^{0}\right) =\sigma _{\gamma }\left( \eta \right) =0.
\end{equation}
On the other hand, $J_{\mu }^{em}$ is invariant (\emph{i.e. }U-spin
symmetric) under the interchange $d\leftrightarrow s,$ which transforms
charged mesons $\pi ^{+}$ and $K^{+}$ into each other: 
\begin{equation}
\sigma _{\gamma }\left( \pi ^{+}\right) =\sigma _{\gamma }\left(
K^{+}\right) \equiv \mu _{\gamma }^{3}
\end{equation}
Consequently, we obtain the generalization of (\ref{psmass}) as 
\begin{eqnarray}
f^{2}m^{2}\left( \pi ^{+}\right)  &=&\left( m_{u}+m_{d}\right) \mu ^{3}+\mu
_{\gamma }^{3}  \nonumber \\
f^{2}m^{2}\left( \pi ^{0}\right)  &=&\left( m_{u}+m_{d}\right) \mu ^{3} 
\nonumber \\
f^{2}m^{2}\left( K^{+}\right)  &=&\left( m_{u}+m_{s}\right) \mu ^{3}+\mu
_{\gamma }^{3} \\
f^{2}m^{2}\left( K^{0}\right)  &=&\left( m_{d}+m_{s}\right) \mu ^{3} 
\nonumber \\
f^{2}m^{2}\left( \eta \right)  &=&\frac{1}{3}\left(
m_{u}+m_{d}+4m_{s}\right) \mu ^{3}.  \nonumber
\end{eqnarray}
From this we can obtain the current quark mass ratios: 
\begin{eqnarray}
\frac{m^{2}\left( K^{0}\right) +\left[ m^{2}\left( K^{+}\right) -m^{2}\left(
\pi ^{+}\right) \right] }{m^{2}\left( K^{0}\right) -\left[ m^{2}\left(
K^{+}\right) -m^{2}\left( \pi ^{+}\right) \right] } &=&\frac{m_{s}}{m_{d}}%
\simeq 20.1  \label{sd-ratio-ps} \\
&&  \nonumber \\
\frac{m^{2}\left( K^{0}\right) -\left[ m^{2}\left( K^{+}\right) -m^{2}\left(
\pi ^{+}\right) \right] }{\left[ 2m^{2}\left( \pi ^{0}\right) -m^{2}\left(
K^{0}\right) \right] +\left[ m^{2}\left( K^{+}\right) -m^{2}\left( \pi
^{+}\right) \right] } &=&\frac{m_{d}}{m_{u}}\simeq 1.8  \label{du-ratio-ps}
\end{eqnarray}
If we assume, for example, $m_{s}\simeq 190\,MeV,$ these ratios yield: 
\begin{equation}
m_{u}\simeq 5.3\,MeV\;\;\;\;\;m_{s}\simeq 9.5\,MeV\;\;\text{or}%
\;\;m_{n}=7.4\,MeV,
\end{equation}
which are indeed very small on the intrinsic scale of QCD. This explains why
the chiral SU(2) and isospin symmetries are such good approximations of the
strong interaction.

\subsection{Quark masses from fitting baryon masses}

For the baryon mass we need to study the matrix elements $\left\langle
B\left| \mathcal{H}\right| B\right\rangle .\;$The flavor $SU\left( 3\right) $
symmetry breaking being given by the quark masses (\ref{Hm}), we need to
evaluate the matrix elements of the quark scalar densities $u_{a}$ between
baryon states : 
\begin{eqnarray*}
\mathcal{H}_{m} &=&m_{u}\bar{u}u+m_{d}\bar{d}d+m_{s}\bar{s}s \\
&=&m_{0}u_{0}+m_{3}u_{3}+m_{8}u_{8}
\end{eqnarray*}
with 
\begin{equation}
\begin{array}{cc}
m_{0}=\frac{1}{3}\left( m_{u}+m_{d}+m_{s}\right) \;\;\; & u_{0}=\bar{u}u+%
\bar{d}d+\bar{s}s \\ 
m_{3}=\frac{1}{2}\left( m_{u}-m_{d}\right) \;\;\; & u_{3}=\bar{u}u-\bar{d}d
\\ 
m_{8}=\frac{1}{6}\left( m_{u}+m_{d}-2m_{s}\right) \;\;\; & u_{8}=\bar{u}u+%
\bar{d}d-2\bar{s}s
\end{array}
\end{equation}
where, instead of the standard $u_{a}=\bar{q}\lambda _{a}q$ ($\lambda _{a}$
being the familiar Gell-Mann matrices), we have used, for our purpose, the
more convenient definitions of scalar densities by moving some numerical
factors into the quark mass combinations $m_{0,3,8}.$

We shall first concentrate on the low lying baryon octet which, being the
adjoint representation of $SU\left( 3\right) ,$ can be written as a $3\times
3$ matrix 
\begin{equation}
\hat{B}=\left( 
\begin{array}{ccc}
\sqrt{\frac{1}{2}}\Sigma ^{0}+\sqrt{\frac{1}{6}}\Lambda & \Sigma ^{+} & p \\ 
\Sigma ^{-} & -\sqrt{\frac{1}{2}}\Sigma ^{0}+\sqrt{\frac{1}{6}}\Lambda & n
\\ 
\Xi ^{-} & \Xi ^{0} & \sqrt{\frac{2}{3}}\Lambda
\end{array}
\right) .  \label{B-matrix}
\end{equation}
The octet scalar densities $u_{a}$ can be related to two parameters
(Wigner-Eckart theorem): 
\[
\left\langle B\left| u_{a}\right| B\right\rangle =\alpha \,tr\left( \hat{B}%
^{\dagger }\hat{u}_{a}\hat{B}\right) +\beta \,tr\left( \hat{B}^{\dagger }%
\hat{B}\hat{u}_{a}\right) 
\]
where $\hat{u}_{a}$ is the scalar density expressed as a $3\times 3$ matrix
in the quark flavor space. The linear combinations $\left( \alpha \pm \beta
\right) /2$ are the familiar $D$ and $F$ coefficients. For example, we can
easily compute: 
\begin{eqnarray}
\left\langle p\left| u_{8}\right| p\right\rangle &=&\alpha -2\beta =\left(
3F-D\right) _{mass}  \label{u8} \\
\left\langle p\left| u_{3}\right| p\right\rangle &=&\alpha =\left(
F+D\right) _{mass}.  \label{u3}
\end{eqnarray}
In this way the baryon masses with their electromagnetic self-energy
subtracted (as denoted by the baryon names) can be expressed in terms of
three parameters 
\begin{eqnarray}
p\; &=&\mathcal{M}_{0}+(\alpha -2\beta )m_{8}+\alpha m_{3}  \label{bary-para}
\\
n\; &=&\mathcal{M}_{0}+(\alpha -2\beta )m_{8}-\alpha m_{3}  \nonumber \\
\Sigma ^{+} &=&\mathcal{M}_{0}+(\alpha +\beta )m_{8}+\left( \alpha -\beta
\right) m_{3}  \nonumber \\
\Sigma ^{0} &=&\mathcal{M}_{0}+(\alpha +\beta )m_{8}  \nonumber \\
\Sigma ^{-} &=&\mathcal{M}_{0}+(\alpha +\beta )m_{8}-\left( \alpha -\beta
\right) m_{3}  \nonumber \\
\Xi ^{-} &=&\mathcal{M}_{0}+(\beta -2\alpha )m_{8}+\beta m_{3}  \nonumber \\
\Xi ^{0} &=&\mathcal{M}_{0}+(\beta -2\alpha )m_{8}-\beta m_{3}  \nonumber \\
\Lambda \; &=&\mathcal{M}_{0}-(\alpha +\beta )m_{8}  \nonumber
\end{eqnarray}
We have 8 baryon masses and three unknown parameters $\mathcal{M}%
_{0},\,\alpha $ and $\beta $ --- hence 5 relations, one of them should yield
quark mass ratio $m_{8}/m_{3}$.

\begin{itemize}
\item  The (``improved'') Gell-Mann-Okubo mass relation 
\begin{equation}
n+\Xi ^{-}=\frac{1}{2}\left( 3\Lambda +2\Sigma ^{+}-\Sigma ^{0}\right) 
\label{GMO-b}
\end{equation}

\item  The Coleman-Glashow (U-spin) relation 
\begin{equation}
\Xi ^{-}-\Xi ^{0}=\left( p-n\right) +\left( \Sigma ^{-}-\Sigma ^{+}\right) 
\label{CG-b}
\end{equation}

\item  Absence of isospin $I=2\;$correction (\emph{i.e.} $u_{3}$ being a
member of $I=1$): 
\begin{equation}
\Sigma ^{-}+\Sigma ^{+}-2\Sigma ^{0}=0  \label{I2-b}
\end{equation}

\item  The hybrid relation: 
\begin{equation}
\frac{p-n}{\Sigma ^{-}-\Xi ^{-}}=\frac{\Xi ^{-}-\Xi ^{0}}{\Sigma ^{+}-p}
\label{hybrid-b}
\end{equation}
It should not be surprising that we have a relation relating $SU\left(
2\right) $ breakings to $SU\left( 3\right) $ breakings, since $u_{3}$ and $%
u_{8}$ belong to the same octet representation. Recall that here the
electromagnetic contribution must be subtracted from our masses (sometimes
called the \emph{tadpole masses}). Since there is no Dashen theorem for the
electromagnetic contributions to baryon masses, we must resort to detailed
(\& less reliable) model calculations. We quote one such result\cite{col-sch}
for the electromagnetic contributions $\left( \Delta M\right) _{\gamma }:$%
\begin{eqnarray*}
p-n=\left( p-n\right) _{obs}-\left( p-n\right) _{\gamma }\simeq
-1.3-1.1\simeq -2.4\,MeV \\
\Xi ^{-}-\Xi ^{0}=\left( \Xi ^{-}-\Xi ^{0}\right) _{obs}-\left( \Xi ^{-}-\Xi
^{0}\right) _{\gamma }\simeq 6.4-1.3\simeq 5.1\,MeV
\end{eqnarray*}
which yields $\simeq 0.02$ on both sides of Eqn.(\ref{hybrid-b}).

\item  Both sides of Eq.(\ref{hybrid-b}) are related to the quark mass ratio 
$2m_{3}/\left( 3m_{8}-m_{3}\right) .\;$Thus the above result leads to 
\begin{equation}
\left( \frac{m_{u}-m_{d}}{m_{d}-m_{s}}\right) _{B}\simeq 0.02
\label{qm-ratio-b}
\end{equation}
which is compatible with the current quark ratio deduced from pseudoscalar
meson masses Eqs.(\ref{sd-ratio-ps}) and (\ref{du-ratio-ps}): 
\begin{equation}
\left( \frac{m_{u}-m_{d}}{m_{d}-m_{s}}\right) _{ps}\simeq \frac{\frac{1}{1.8}%
-1}{1-20.1}\simeq 0.023.  \label{qmass-b}
\end{equation}
\end{itemize}

\subsection{The constituent quark model}

\subsubsection{Spin-dependent contributions to baryon masses}

The sQM which attempts to describe the properties of light hadrons as a
composite systems of $u,d,$ and $s$ valence quarks. The mass relations
derived above may be interpreted simply as reflecting the hadrons masses as
sum of the corresponding valence quark masses. For a general baryon, we have 
\begin{equation}
B=\mathcal{M}_{0}+M_{1}+M_{2}+M_{3}  \label{baryon3m}
\end{equation}
where $\mathcal{M}_{0}$ is some $SU\left( 3\right) $ symmetric binding
contribution. $M_{1,2,3}$ are the \emph{constituent masses }of the three
valence quarks. We shall ignore isospin breaking effects: $M_{u}=M_{d}\equiv
M_{n}$, and write the octet baryon masses as, 
\begin{eqnarray}
N &=&\mathcal{M}_{0}+3M_{n}  \label{b-oct-3m} \\
\Lambda  &=&\mathcal{M}_{0}+2M_{n}+M_{s}  \nonumber \\
\Sigma  &=&\mathcal{M}_{0}+2M_{n}+M_{s}  \nonumber \\
\Xi  &=&\mathcal{M}_{0}+M_{n}+2M_{s},  \nonumber
\end{eqnarray}
and the decuplet baryon masses as, 
\begin{eqnarray}
\Delta  &=&\mathcal{M}_{0}+3M_{n}  \label{b-dec-3m} \\
\Sigma ^{*} &=&\mathcal{M}_{0}+2M_{n}+M_{s}  \nonumber \\
\Xi ^{*} &=&\mathcal{M}_{0}+M_{n}+2M_{s}  \nonumber \\
\Omega  &=&\mathcal{M}_{0}+3M_{s}.  \nonumber
\end{eqnarray}
While it reproduces the GMO mass relations respectively, for the octet: 
\begin{equation}
N+\Xi =\frac{1}{2}(3\Lambda +\Sigma ),  \label{GMO-b1}
\end{equation}
and for the decuplet (the equal-spacing rule): 
\begin{equation}
\Delta -\Sigma ^{*}=\Sigma ^{*}-\Xi ^{*}=\Xi ^{*}-\Omega ,  \label{GMO-b2}
\end{equation}
it also leads to a phenomenologically incorrect result of $\Lambda =\Sigma $
(reflecting their identical quark contents). Similarly, such a naive picture
would lead us to expect that the $N,\,\Sigma ,\,\Xi $ baryons having
comparable masses as $\Delta ,\,\Sigma ^{*},\,\Xi ^{*}$. Observationally the
spin 3/2 decuplet has significantly higher masses than the spin 1/2 octet
baryons. Similar pattern has also been observed in the meson spectrum: the
spin 1 meson octet is seen to be significantly heavier than the spin 0
mesons: $M_{\rho ,K^{*},\omega }\gg M_{\pi ,K,\eta }$ even though they have
the same quark contents. This suggests that there must be important
spin-dependent contributions to these light hadron masses\cite{DRGG}. We
then generalize Eq.(\ref{baryon3m}) to 
\begin{equation}
B=\mathcal{M}_{0}+M_{1}+M_{2}+M_{3}+\kappa \left[ \left( \frac{\mathbf{s}_{1}%
\mathbf{\cdot s}_{2}}{M_{1}M_{2}}\right) +\left( \frac{\mathbf{s}_{3}\mathbf{%
\cdot s}_{2}}{M_{3}M_{2}}\right) +\left( \frac{\mathbf{s}_{1}\mathbf{\cdot s}%
_{3}}{M_{1}M_{3}}\right) \right]   \label{bmass-spin}
\end{equation}
where $\mathbf{s}_{i}\;$is the spin of i-th quark, and the constant $\kappa $
one would adjust to fit the experimental data. This spin dependent
contribution is modeled after the hyperfine splitting of atomic physics. For
hydrogen atom we have a two body system hence only one pair of spin-spin
interaction: $M_{1}=m_{e}$ and $M_{2}=M_{p}.$ The $\left( \frac{\mathbf{s}%
_{e}\mathbf{\cdot s}_{p}}{m_{e}M_{p}}\right) $ arises from $\mathbf{\mu
\cdot B\sim \mu }_{e}\mathbf{\cdot \mu }_{p}/r^{3}\mathbf{\ }$with the
proportional constant worked out to be 
\[
\kappa _{H}=\frac{8\pi e^{2}\mu _{p}}{3}\left| \psi \left( 0\right) \right|
^{2}
\]
where $\mu _{p}=2.79\,$is the magnetic moment of the proton in unit of
nucleon magneton, and $\psi \left( 0\right) $ is the hydrogen wave function
at origin. Such an interaction accounts for the $1420\,MHz$ splitting
between the two 1S states, which gives rise to the famous $21$\thinspace $%
cm\;$line of hydrogen. For the case of baryon, one usually attributes such
interaction to one-gluon exchange; but we shall comment on this point in
later part of these lectures, at the end of Sec. 3.2.

To compute the $\frac{\mathbf{s}_{i}\mathbf{\cdot s}_{j}}{M_{i}M_{j}}$ terms
we need to distinguish three cases:

(a) The equal mass case: $M_{1}=M_{2}=M_{3}\equiv M$%
\begin{eqnarray}
&&\left[ \left( \frac{\mathbf{s}_{1}\mathbf{\cdot s}_{2}}{M_{1}M_{2}}\right)
+\left( \frac{\mathbf{s}_{1}\mathbf{\cdot s}_{2}}{M_{1}M_{2}}\right) +\left( 
\frac{\mathbf{s}_{1}\mathbf{\cdot s}_{2}}{M_{1}M_{2}}\right) \right] =\frac{1%
}{M^{2}}\left( \sum_{i>j}\mathbf{s}_{i}\cdot \mathbf{s}_{j}\right)  \nonumber
\\
&=&\frac{1}{2M^{2}}\left( \mathbf{S}^{2}-\mathbf{s}_{1}^{2}-\mathbf{s}%
_{2}^{2}-\mathbf{s}_{3}^{2}\right) =\frac{1}{2M^{2}}\left[ S\left(
S+1\right) -3s\left( s+1\right) \right]  \nonumber \\
&=&\QATOPD\{ . {-\frac{3}{4M^{2}}\;\;\;\text{for\ \ \ }S=1/2}{+\frac{3}{%
4M^{2}}\;\;\;\text{for\ \ \ }S=3/2}
\end{eqnarray}
This is applicable for the $N,\,\Delta ,\,\Omega $ baryons.

(b) The unequal mass case, for example, $\left( ssn\right) :$ Because of
color antisymmetrization, the baryon wavefunction must be symmetric under
the combined interchange of flavor and spin labels. Since we have a
symmetric superposition of flavor states, the$\,$subsystem $\left( ss\right) 
$ must have spin 1, Namely, $\mathbf{s}_{s}\cdot \mathbf{s}_{s}=\frac{1}{2}%
(2-2\mathbf{s}_{s}^{2})=\frac{1}{4},$ and 
\[
2\mathbf{s}_{s}\cdot \mathbf{s}_{n}=\left( \sum_{i>j}\mathbf{s}_{i}\cdot 
\mathbf{s}_{j}\right) -\mathbf{s}_{s}\cdot \mathbf{s}_{s}=\QATOPD\{ . {-%
\frac{3}{4}-\frac{1}{4}=-1\;\;\;\text{for\ \ \ }S=1/2}{+\frac{3}{4}-\frac{1}{%
4}=+\frac{1}{2}\;\;\;\text{for\ \ \ }S=3/2} 
\]
or 
\begin{eqnarray}
&&\left[ \left( \frac{\mathbf{s}_{1}\mathbf{\cdot s}_{2}}{M_{1}M_{2}}\right)
+\left( \frac{\mathbf{s}_{1}\mathbf{\cdot s}_{2}}{M_{1}M_{2}}\right) +\left( 
\frac{\mathbf{s}_{1}\mathbf{\cdot s}_{2}}{M_{1}M_{2}}\right) \right] =\left[
\left( \frac{\mathbf{s}_{s}\mathbf{\cdot s}_{s}}{M_{s}^{2}}\right) +2\left( 
\frac{\mathbf{s}_{s}\mathbf{\cdot s}_{n}}{M_{s}M_{n}}\right) \right] 
\nonumber \\
&=&\QATOPD\{ . {\frac{1}{4M_{s}^{2}}-\frac{1}{M_{s}M_{n}}\;\;\;\text{for\ \
\ }S=1/2}{\frac{1}{4M_{s}^{2}}+\frac{1}{2M_{s}M_{n}}\;\;\;\text{for\ \ \ }%
S=3/2}
\end{eqnarray}
This case is applicable to $\Xi $ and $\Xi ^{*},$ as well as $\Sigma $ and $%
\Sigma ^{*}$ because the sigma baryons are isospin $I=1$ states (hence
symmetric in the nonstrange flavor space).

(c) The $\Lambda $ baryon: Because $\Lambda $ is an isoscalar, the subsystem
must be in spin 0 state. From this one can easily work out the spin factor
to be $-\frac{3}{4M_{n}^{2}},$ independent of $M_{s}.$

Putting all this together into Eq.(\ref{bmass-spin}) we obtain, for the
octet baryons: 
\begin{eqnarray}
N &=&\mathcal{M}_{0}+3M_{n}-\frac{3\kappa }{4M_{n}^{2}} \\
\Lambda &=&\mathcal{M}_{0}+2M_{n}+M_{s}-\frac{3\kappa }{4M_{n}^{2}} 
\nonumber \\
\Sigma &=&\mathcal{M}_{0}+2M_{n}+M_{s}+\frac{\kappa }{4M_{n}^{2}}-\frac{%
\kappa }{M_{s}M_{n}}  \nonumber \\
\Xi &=&\mathcal{M}_{0}+M_{n}+2M_{s}+\frac{\kappa }{4M_{s}^{2}}-\frac{\kappa 
}{M_{s}M_{n}}  \nonumber
\end{eqnarray}
and for the decuplet baryons: 
\begin{eqnarray}
\Delta &=&\mathcal{M}_{0}+3M_{n}+\frac{3\kappa }{4M_{n}^{2}} \\
\Sigma ^{*} &=&\mathcal{M}_{0}+2M_{n}+M_{s}+\frac{\kappa }{4M_{n}^{2}}+\frac{%
\kappa }{2M_{s}M_{n}}  \nonumber \\
\Xi ^{*} &=&\mathcal{M}_{0}+M_{n}+2M_{s}+\frac{\kappa }{4M_{s}^{2}}+\frac{%
\kappa }{2M_{s}M_{n}}  \nonumber \\
\Omega &=&\mathcal{M}_{0}+3M_{s}+\frac{3\kappa }{4M_{s}^{2}}.  \nonumber
\end{eqnarray}
One can obtain an excellent fit (within 1\%) to all the masses with the
parameter values (\emph{e.g. }\cite{rosner}) $\mathcal{M}_{0}=0,\;\frac{%
\kappa }{M_{n}^{2}}=50\,MeV\;$and the constituent quark mass values of 
\begin{equation}
M_{n}=363\;MeV,\;\;\;\;\;M_{s}=538\,MeV.  \label{cq-mass-ss}
\end{equation}

Similarly good fit can also be obtained for mesons, with an enhanced value
of $\kappa .$ Besides some different coupling factors this may reflect a
larger $\left| \psi \left( 0\right) \right| ^{2}\propto R^{-3},$ which is
compatible with the observed root mean square charge radii of mesons \emph{vs%
} baryons: $R_{meson}\simeq 0.6\,fm$ vs $R_{baryon}\simeq 0.8\,fm.$

\subsubsection{Spin and magnetic moments of the baryon}

Another useful tool to study hadron structure is the magnetic moment of the
baryon. Their deviation from the Dirac moments values ($e_{B}/2M_{B}$)
indicates the presence of structure. In the quark model the simplest
possibility is that the baryon magnetic moment is simply the sum of its
constituent quark's Dirac moments. Clearly, the magnetic moments are
intimately connected to the spin structure of the hadron. Hence, we shall
first make a detour into a discussion of the baryon spin structure in the
constituent quark model.

\paragraph{Quark contributions to the proton spin}

Because it is antisymmetric under the interchange of quark color indices,
the baryon wavefunction must be symmetric in the spin-flavor space.
Mathematically, we say that the baryon wavefunction should be invariant
under the permutation group $S_{3}$ --- the group of permuting three quarks
with spin and isospin labels.

We shall concentrate on the case of proton. While the product wavefunction
is symmetric, the individual spin and isospin wavefunctions are of the
mixed-symmetry type. There are two mixed-symmetry spin-$\frac12$
wavefunction combinations:

\begin{quote}
(i) $\chi _{S}$ --- \emph{symmetric in the first two quarks:} Namely, the
first two quarks form a spin 1 subsystem: (Notation for the spin-up and
-down states: $\left| \frac{1}{2},+\frac{1}{2}\right\rangle \equiv \alpha $
and $\left| \frac{1}{2},-\frac{1}{2}\right\rangle \equiv \beta $) 
\[
\left| 1,+1\right\rangle =\alpha _{1}\alpha _{2},\;\;\;\left|
1,0\right\rangle =\frac{1}{\sqrt{2}}\left( \alpha _{1}\beta _{2}+\beta
_{1}\alpha _{2}\right) ,\;\;\;\left| 1,-1\right\rangle =\beta _{1}\beta
_{2}\;\;\; 
\]
which is combined with the 3rd quark to form a spin $\frac{1}{2}$ proton: 
\[
\left| \frac{1}{2},+\frac{1}{2}\right\rangle _{S}=\sqrt{\frac{2}{3}}\left|
1,+1\right\rangle \left| \frac{1}{2},-\frac{1}{2}\right\rangle -\sqrt{\frac{1%
}{3}}\left| 1,0\right\rangle \left| \frac{1}{2},+\frac{1}{2}\right\rangle 
\]
or 
\begin{equation}
\chi _{S}=\frac{1}{\sqrt{6}}(2\alpha _{1}\alpha _{2}\beta _{3}-\alpha
_{1}\beta _{2}\alpha _{3}-\beta _{1}\alpha _{2}\alpha _{3}).
\end{equation}

(ii) $\chi _{A}$ --- \emph{antisymmetric in the first two quarks:} The first
two quarks form a spin 0 subsystem: 
\[
\left| \frac{1}{2},+\frac{1}{2}\right\rangle _{A}=\left| 0,0\right\rangle
\left| \frac{1}{2},+\frac{1}{2}\right\rangle 
\]
or 
\begin{equation}
\chi _{A}=\frac{1}{\sqrt{2}}(\alpha _{1}\beta _{2}-\beta _{1}\alpha
_{2})\alpha _{3}.
\end{equation}
\end{quote}

While $\chi _{S,A}$ are the \emph{spin-$\frac{1}{2}$} wavefunctions, with
identical steps, we can construct the two mixed-symmetry\emph{\ isospin-$%
\frac{1}{2}$} wavefunctions $\chi _{S,A}^{\prime }:$%
\begin{eqnarray}
\chi _{S}^{\prime } &=&\frac{1}{\sqrt{6}}%
(2u_{1}u_{2}d_{3}-u_{1}d_{2}u_{3}-d_{1}u_{2}u_{3})  \nonumber \\
\chi _{A}^{\prime } &=&\frac{1}{\sqrt{2}}(u_{1}d_{2}-d_{1}u_{2})u_{3}.
\end{eqnarray}
Both the spin wavefunctions $\left( \chi _{S}\;\chi _{A}\right) $ and the
isospin wavefunctions $\left( \chi _{S}^{\prime }\;\chi _{A}^{\prime
}\right) $ form a two dimensional representation of the permutation group $%
S_{3}.$ For example, under the permutation operations of $P_{12}$ and $%
P_{13} $%
\[
P_{12}\binom{\chi _{S}}{\chi _{A}}=\stackunder{M_{12}}{\underbrace{\left( 
\begin{array}{cc}
1 & \;0 \\ 
0 & -1
\end{array}
\right) }}\binom{\chi _{S}}{\chi _{A}}\;\;\;\;\;\;\;P_{13}\binom{\chi _{S}}{%
\chi _{A}}=\stackunder{M_{13}}{\underbrace{\left( 
\begin{array}{cc}
-\frac{1}{2} & -\frac{\sqrt{3}}{2} \\ 
-\frac{\sqrt{3}}{2} & +\frac{1}{2}
\end{array}
\right) }}\binom{\chi _{S}}{\chi _{A}} 
\]
where $M_{ij}$ are 2-dimensional representations in terms of orthogonal
matrices. Consequently, we find that the combinations such as $\left( \chi
_{S}^{2}+\chi _{A}^{2}\right) ,\;\left( \chi _{S}^{\prime 2}+\chi
_{A}^{\prime 2}\right) $ and $\left( \chi _{S}\chi _{S}^{\prime }+\chi
_{A}\chi _{A}^{\prime }\right) $ are invariant under $S_{3}$
transformations. In this way we find the symmetric proton spin-isospin
wavefunction: 
\begin{eqnarray}
\left| p_{+}\right\rangle &=&\frac{1}{\sqrt{2}}\left( \chi _{S}\chi
_{S}^{\prime }+\chi _{A}\chi _{A}^{\prime }\right) \\
&=&\frac{1}{\sqrt{2}}[\frac{1}{6}(2\alpha _{1}\alpha _{2}\beta _{3}-\alpha
_{1}\beta _{2}\alpha _{3}-\beta _{1}\alpha _{2}\alpha
_{3})(2u_{1}u_{2}d_{3}-u_{1}d_{2}u_{3}-d_{1}u_{2}u_{3})  \nonumber \\
&&+\frac{1}{2}(\alpha _{1}\beta _{2}\alpha _{3}-\beta _{1}\alpha _{2}\alpha
_{3})(u_{1}d_{2}u_{3}-d_{1}u_{2}u_{3})]  \nonumber \\
&=&\frac{1}{6\sqrt{2}}[4\left(
u_{+}u_{+}d_{-}+u_{+}d_{-}u_{+}+d_{-}u_{+}u_{+}\right)  \nonumber \\
&&\;\;\;\;\;-2(u_{+}u_{-}d_{+}+u_{-}d_{+}u_{+}+d_{+}u_{+}u_{-}  \nonumber \\
&&\;\;\;\;\;\;+u_{-}u_{+}d_{+}+u_{+}d_{+}u_{-}+d_{+}u_{-}u_{+})]  \nonumber
\end{eqnarray}
where we have used the notation of $\alpha u=u_{+},\;\beta d=d_{-},\,etc.\;$%
In calculating physical quantities, many terms, \emph{e.g. }$%
u_{+}u_{+}d_{-},\;u_{+}d_{-}u_{+}$ and $d_{-}u_{+}u_{+}$ yield the same
contribution. Hence we can use the simplified wavefunction: 
\begin{equation}
\left| p_{+}\right\rangle =\frac{1}{\sqrt{6}}%
(2u_{+}u_{+}d_{-}-u_{+}u_{-}d_{+}-u_{-}u_{+}d_{+})  \label{spinWF}
\end{equation}
From this we can count the \emph{number of quark flavors} with spin parallel
or antiparallel to the proton spin: 
\begin{equation}
u_{+}=\frac{5}{3},\;\;u_{-}=\frac{1}{3},\;\;\;d_{+}=\frac{1}{3},\;\;u_{-}=%
\frac{2}{3}  \label{spin-density}
\end{equation}
summing up to two $u$ and one $d$ quarks. From the difference 
\begin{equation}
\Delta q=q_{+}-q_{-}  \label{del-q}
\end{equation}
we also obtain the contribution by each of the quark flavors to the proton
spin: 
\begin{equation}
\Delta u=\frac{4}{3}\;\;\;\;\;\Delta d=-\frac{1}{3}\;\;\;\;\;\Delta
s=0,\;\;\;\;\text{and \ \ }\Delta \Sigma =1,  \label{sqm-delq}
\end{equation}
where $\Delta \Sigma =\Delta u+\Delta d+\Delta s$ is the sum of quark
polarizations.

\paragraph{Quark contributions to the baryon magnetic moments}

Instead of proceeding directly to the results of quark model calculation of
the baryon magnetic moments, we shall first set up a more general framework.
This will be useful when we consider the contribution from the quark sea in
the later part of these lectures. We shall pay special attention to the
contribution by antiquarks. If there are antiquarks in the proton, the
definition in Eq.(\ref{del-q}) becomes 
\begin{equation}
\Delta q=\left( q_{+}-q_{-}\right) +\left( \bar{q}_{+}-\bar{q}_{-}\right)
\equiv \Delta _{q}+\Delta _{\overline{q}}  \label{del-qqbar}
\end{equation}
Thus the quark spin contribution $\Delta q$ is the \emph{sum }of the quark
and antiquark polarizations. For the $q$-flavor quark contribution to the
proton magnetic moment, we have however 
\begin{equation}
\mu _{p}\left( q\right) =\Delta _{q}\mu _{q}+\Delta _{\overline{q}}\mu _{%
\overline{q}}=\left( \Delta _{q}-\Delta _{\overline{q}}\right) \mu
_{q}\equiv \widetilde{\Delta q}\mu _{q}  \label{deldiff}
\end{equation}
where $\mu _{q}$ is the magnetic moment of the $q$-flavor quark. The
negative sign simply reflects the opposite quark and antiquark moments, $\mu
_{\overline{q}}=-\mu _{q}.$ Thus the spin factor that enters into the
expression for the magnetic moment is $\widetilde{\Delta q}$, the \emph{%
difference} of the quark and antiquark polarizations. If we assume that the
proton magnetic moment is entirely built up from the light quarks inside it,
we have 
\begin{equation}
\mu _{p}=\widetilde{\Delta u}\mu _{u}+\widetilde{\Delta d}\mu _{d}+%
\widetilde{\Delta s}\mu _{s}.  \label{proton}
\end{equation}
In such an expression there is a \emph{separation} of the intrinsic quark
magnetic moments and the spin wavefunctions. Flavor-$SU(3)$ symmetry then
implies, the proton wavefunction being related the $\Sigma ^{+}$
wavefunction by the interchange of $d\leftrightarrow s$ and $\overline{d}%
\leftrightarrow \overline{s}\;$quarks, the relations $\left( \widetilde{%
\Delta u}\right) _{\Sigma ^{+}}=\left( \widetilde{\Delta u}\right)
_{p}\equiv \widetilde{\Delta u},$ $\left( \widetilde{\Delta d}\right)
_{\Sigma ^{+}}=\widetilde{\Delta s},$ and$\;\left( \widetilde{\Delta s}%
\right) _{\Sigma ^{+}}=\widetilde{\Delta d};$ similarly it being related to
the $\Xi ^{0}$ wavefunction by a further interchange of $u\leftrightarrow s$
quarks, thus $\left( \widetilde{\Delta d}\right) _{\Xi ^{0}}=\left( 
\widetilde{\Delta d}\right) _{\Sigma ^{+}}=\widetilde{\Delta s},\;\left( 
\widetilde{\Delta s}\right) _{\Xi ^{0}}=\left( \widetilde{\Delta u}\right)
_{\Sigma ^{+}}=\widetilde{\Delta u},$ and $\left( \widetilde{\Delta u}%
\right) _{\Xi ^{0}}=\left( \widetilde{\Delta s}\right) _{\Sigma ^{+}}=%
\widetilde{\Delta d}.$ We have, 
\begin{eqnarray}
\mu _{\Sigma ^{+}} &=&\widetilde{\Delta u}\mu _{u}+\widetilde{\Delta s}\mu
_{d}+\widetilde{\Delta d}\mu _{s},  \label{sigma+} \\
\mu _{\Xi ^{0}} &=&\widetilde{\Delta d}\mu _{u}+\widetilde{\Delta s}\mu _{d}+%
\widetilde{\Delta u}\mu _{s},  \label{cascade0}
\end{eqnarray}
the intrinsic moments $\mu _{q}$ being unchanged when we go from Eq.(\ref
{proton}) to Eqs.(\ref{sigma+}) and (\ref{cascade0}). The $n,\;\Sigma ^{-}$,
and $\Xi ^{-}$ moments can be obtained from their isospin conjugate partners 
$p,\Sigma ^{+}$, and $\Xi ^{0}$ by the interchange of their respective $%
u\leftrightarrow d$ quarks: $\left( \widetilde{\Delta u}\right) _{\Sigma
^{-}}=\left( \widetilde{\Delta d}\right) _{\Sigma ^{+}}=\widetilde{\Delta s}%
,\;etc.$%
\begin{eqnarray}
\mu _{n} &=&\widetilde{\Delta d}\mu _{u}+\widetilde{\Delta u}\mu _{d}+%
\widetilde{\Delta s}\mu _{s},  \label{neutron} \\
\mu _{\Sigma ^{-}} &=&\widetilde{\Delta s}\mu _{u}+\widetilde{\Delta u}\mu
_{d}+\widetilde{\Delta d}\mu _{s},  \label{sigma-} \\
\mu _{\Xi ^{-}} &=&\widetilde{\Delta s}\mu _{u}+\widetilde{\Delta d}\mu _{d}+%
\widetilde{\Delta u}\mu _{s}.  \label{cascade-}
\end{eqnarray}
The relations for the $I_{z}=0,\;Y=0$ moments are more complicated in
appearance but the underlying arguments are the same.

\begin{eqnarray}
\mu _{\Lambda } &=&\frac{1}{6}\left( \widetilde{\Delta u}+4\widetilde{\Delta
d}+\widetilde{\Delta s}\right) \left( \mu _{u}+\mu _{d}\right)  \label{lamda}
\\
&&\ \ +\frac{1}{6}\left( 4\widetilde{\Delta u}-2\widetilde{\Delta d}+4%
\widetilde{\Delta s}\right) \mu _{s},  \nonumber \\
\mu _{\Lambda \Sigma } &=&\frac{-1}{2\sqrt{3}}\left( \widetilde{\Delta u}-2%
\widetilde{\Delta d}+\widetilde{\Delta s}\right) \left( \mu _{u}-\mu
_{d}\right) .  \label{transmom}
\end{eqnarray}

In the nonrelativistic constituent quark model, there is no quark sea and
hence no antiquark polarization, $\Delta _{\overline{q}}=0.$ This means that
in the sQM we have $\Delta q=\widetilde{\Delta q}.\;$After plugging in the
result of Eq.(\ref{sqm-delq}), we obtain the result in the 2nd column of the
Table 1 below:

\[
\begin{tabular}{|ccccc|}
\hline
Baryon & mag moment & $u=-2d$ & $d=-0.9\mu _{N}$ & exptl \# \\ 
& $\left( q\equiv \mu _{q}\right) $ & $s=2d/3$ &  & $\left( \mu _{N}\right) $
\\ \hline
&  &  &  &  \\ 
$p$ & $\left( 4u-d\right) /3$ & $-3d$ & $2.7$ & $\,2.79$ \\ 
&  &  &  &  \\ 
$n$ & $\left( 4d-u\right) /3$ & $2d$ & $-1.8$ & $-1.91$ \\ 
&  &  &  &  \\ 
$\Sigma ^{+}$ & $\left( 4u-s\right) /3$ & $-26d/9$ & $2.6$ & $2.48$ \\ 
&  &  &  &  \\ 
$\Sigma ^{-}$ & $\left( 4d-s\right) /3$ & $10d/9$ & $-1.0$ & $-1.16$ \\ 
&  &  &  &  \\ 
$\Xi ^{0}$ & $\left( 4s-u\right) /3$ & $14d/9$ & $-1.4$ & $-1.25$ \\ 
&  &  &  &  \\ 
$\Xi ^{-}$ & $\left( 4s-d\right) /3$ & $5d/9$ & $-0.5$ & $-0.68$ \\ 
&  &  &  &  \\ 
$\Lambda $ & $s$ & $2d/3$ & $-0.6$ & $-0.61$ \\ 
&  &  &  &  \\ 
$\Lambda \Sigma $ & $\left( d-u\right) /\sqrt{3}$ & $\sqrt{3}d$ & $-1.6$ & $%
-1.60$ \\ \hline
\end{tabular}
\]

\begin{center}
\underline{Table 1}. Quark contribution to the octet baryon magnetic
moments.\medskip
\end{center}

Instead of trying to get the best fit at this stage, we shall simplify the
result further with the following observation: Because of the assumption $%
M_{u}=M_{d},$ we have $\mu _{u}=-2\mu _{d}.$ The proton and neutron moments
are then reduced to $\mu _{p}=-3\mu _{d}$ and $\mu _{n}=2\mu _{d},$ and thus
the ratio 
\begin{equation}
\frac{\mu _{p}}{\mu _{n}}=-1.5
\end{equation}
which is very close to the experimental value of $-1.48.$ Furthermore, we
have seen in previous discussion that constituent strange-quark mass is
about a third heavier than the nonstrange quarks $M_{s}/M_{n}\simeq 3/2$, we
can make the approximation of $\mu _{s}=2\mu _{d}/3.$ In this way, all the
moments are expressed in terms of the $d$ quark moment, as displayed in the
3rd column above. One can then make a best over-all-fit to the experimental
values by adjusting this last parameter $\mu _{d}.\;$The final results, in
column 4, are obtained by taking $\mu _{d}=-0.9\,\mu _{N}$, where $\mu _{N}$
is nucleon magneton $e/2M_{N}$. They are compared, quite favorably, with the
experimental values in the last column. We also note that, with the $d$
quark having a third of the electronic charge, the fit-parameter of $\mu
_{d}=-0.9\,\mu _{N}$ translates into a $d$ quark constituent mass of 
\begin{equation}
M_{n}=\frac{M_{N}}{3\times 0.9}=348\,MeV,\;\;\text{and\ \ \ }M_{s}=\frac{%
3M_{n}}{2}=522\,MeV,  \label{cq-mass-mm}
\end{equation}
which are entirely compatible with the constituent quark mass values in Eq.(%
\ref{cq-mass-ss}), obtained in fitting the baryon masses by including the
spin-dependent contributions.

\subsubsection{sQM lacks a quark sea}

So far we have discussed the successes of the simple quark model. There are
several instances which indicate that this model is too simple: sQM does not
yield the correct nucleon matrix elements of the axial vector and scalar
density operators.

\paragraph{Axial vector current matrix elements}

The quark spin contribution to proton $\Delta q$ in Eq.(\ref{del-qqbar}) is
just the proton matrix element of the quark axial vector current operator 
\begin{equation}
2s_{\mu }\Delta q=\left\langle p,s\left| \bar{q}\gamma _{\mu }\gamma
_{5}q\right| p,s\right\rangle =2s_{\mu }\left( q_{+}-q_{-}+\bar{q}_{+}-\bar{q%
}_{-}\right)  \label{delq-def}
\end{equation}
where $s_{\mu }$ is the spin-vector of the nucleon, as the axial current
vector corresponds to the non-relativistic spin operator: 
\begin{equation}
\bar{q}\mathbf{\gamma }\gamma _{5}q=q^{\dagger }\left( 
\begin{array}{cc}
\mathbf{\sigma } & 0 \\ 
0 & \mathbf{\sigma }
\end{array}
\right) q.  \label{axial-spin}
\end{equation}
Through $SU\left( 3\right) \;$these matrix elements can be related to the
axial vector coupling as measured in the octet baryon beta decays. In
particular, we have 
\begin{eqnarray}
\left( \Delta u-\Delta d\right) _{\text{exptl}} &=&1.26  \nonumber \\
\left( \Delta u+\Delta d-2\Delta s\right) _{\text{expt}} &=&0.6
\label{delq38-exp}
\end{eqnarray}
which is to be compared to the sQM results of Eq.(\ref{sqm-delq}): 
\begin{eqnarray}
\left( \Delta u-\Delta d\right) _{\text{sQM}} &=&5/3  \nonumber \\
\left( \Delta u+\Delta d-2\Delta s\right) _{\text{sQM}} &=&1.
\label{delq38-sqm}
\end{eqnarray}

\paragraph{Scalar density matrix elements}

The matrix elements of scalar density operator $\bar{q}q$ can be interpreted
as number counts of a quark flavor in proton 
\begin{equation}
\left\langle p\left| \bar{q}q\right| p\right\rangle =q+\bar{q}
\end{equation}
where $q\,\left( \bar{q}\right) $ on the RHS denotes the number of a quark
(antiquark) flavor in a proton. Namely, the proton matrix element of the
scalar operator $\bar{q}q$ measures the sum of quark and antiquark number in
the proton (opposed to the difference $q-\bar{q}$ as measured by $q^{\dagger
}q$ ). It is useful to define the \emph{fraction of a quark-flavor} in a
proton as 
\begin{equation}
F\left( q\right) =\frac{\left\langle p\left| \bar{q}q\right| p\right\rangle 
}{\left\langle p\left| \bar{u}u+\bar{d}d+\bar{s}s\right| p\right\rangle }.
\label{frac-def}
\end{equation}
We already have calculated proton matrix element of the scalar density in
the subsection on the baryon masses, Eqs.(\ref{u3}) and (\ref{u8}). Thus we
have 
\begin{equation}
\frac{F\left( 3\right) }{F\left( 8\right) }=\frac{F\left( u\right) -F\left(
d\right) }{F\left( u\right) +F\left( d\right) -2F\left( s\right) }=\frac{%
\alpha }{\alpha -2\beta }  \label{frac38}
\end{equation}
The parameters $\left( \alpha ,\,\beta \right) $ can be deduced from Eq.(\ref
{bary-para}) in the $SU\left( 3\right) $ symmetric limit $\left(
m_{3}u_{3}=0\right) ,\;$as 
\[
\alpha =\frac{M_{\Sigma }-M_{\Xi }}{3m_{8}}\;\;\;\;\;\beta =\frac{M_{\Sigma
}-M_{N}}{3m_{8}} 
\]
Thus the ratio 
\begin{equation}
\left[ \frac{F\left( 3\right) }{F\left( 8\right) }\right] _{\text{exptl}}=%
\frac{M_{\Sigma }-M_{\Xi }}{2M_{N}-M_{\Sigma }-M_{\Xi }}=0.23
\label{f3f8-exp}
\end{equation}
which is to be compared to the sQM value of 
\begin{equation}
\left[ \frac{F\left( 3\right) }{F\left( 8\right) }\right] _{\text{sQM}}=%
\frac{1}{3}.  \label{f3f8-sqm}
\end{equation}
The simplest interpretation of these failures is that the sQM lacks a quark
sea. Hence the number counts of the quark flavors does not come out
correctly.

\subsection{The OZI rule}

The simple quark model of hadron structure discussed above ignores the
presence of quark sea. Even when the issue of the quark sea in nonstrange
hadrons is discussed, its $\left( s\bar{s}\right) $ component is usually
assumed to be highly suppressed$.$ This is based on the OZI-rule\cite{OZI},
which was first deduced from meson mass spectra. In this Subsection we
briefly review this topic.

\subsubsection{The OZI rule for mesons}

The three $\left( q\bar{q}\right) $ combinations that are diagonal in
light-quark flavors are the two isospin $I=1$ and $0$ states of a
flavor-SU(3) octet together with a SU(3) singlet. Isospin being a good
flavor symmetry, there should be very little mixing between the $I=1$ and $0$
states. On the other hand, the flavor-SU(3) being not as a good symmetry, we
anticipate some mixing between the octet- and the singlet- $I=0\;$states: 
\begin{equation}
\left| 8\right\rangle =\frac{1}{\sqrt{6}}\left( u\bar{u}+d\bar{d}-2s\bar{s}%
\right) \;\;\;\;\;\;\;\;\;\;\left| 0\right\rangle =\frac{1}{\sqrt{3}}\left( u%
\bar{u}+d\bar{d}+s\bar{s}\right)  \label{8and0-combs}
\end{equation}

\paragraph{Pseudoscalar meson masses and mixings}

The Gell-Mann-Okubo mass relation for the 0$^{-}$ mesons, before the
identification of $\eta $ as the 8th member of the octet, may be interpreted
as giving the mass of this 8th meson: 
\begin{equation}
m_{8}^{2}=\frac{1}{3}\left( 4m_{K}^{2}-m_{\pi }^{2}\right) =\left(
567\,MeV\right) ^{2}  \label{psGMO}
\end{equation}
which is much closer to the $\eta $ meson mass of $m_{\eta }=547\,MeV$ than $%
m_{\eta ^{\prime }}=958\,MeV.\;$The small difference $m_{8}-m_{\eta }$ can
be attributed to a slight mixing between the octet and singlet isoscalars.
Namely, we interpret $\eta $ and $\eta ^{\prime }$ mesons as two orthogonal
combinations of $\left| 8\right\rangle $ and $\left| 0\right\rangle $ with a
mixing angle that can be determined as follows: 
\[
\left( 
\begin{array}{cc}
m_{8}^{2} & m_{08}^{2} \\ 
m_{80}^{2} & m_{0}^{2}
\end{array}
\right) =\left( 
\begin{array}{cc}
\;\cos \theta & \sin \theta \\ 
-\sin \theta & \cos \theta
\end{array}
\right) \left( 
\begin{array}{cc}
m_{\eta }^{2} & 0 \\ 
0 & m_{\eta ^{\prime }}^{2}
\end{array}
\right) \left( 
\begin{array}{cc}
\cos \theta & -\sin \theta \\ 
\sin \theta & \;\cos \theta
\end{array}
\right) . 
\]
Hence 
\begin{equation}
\sin ^{2}\theta _{P}=\frac{m_{8}^{2}-m_{\eta }^{2}}{m_{\eta ^{\prime
}}^{2}-m_{\eta }^{2}}\;\;\;\;\;\;\;\;\emph{i.e.}\text{\ \ a small \ }\theta
_{P}\simeq 11^{o}.  \label{ps-mix}
\end{equation}

\paragraph{Vector meson masses and mixings}

We now apply the same calculation to the case of vector mesons: 
\[
m_{8}^{*2}=\frac{1}{3}\left( 4m_{K*}^{2}-m_{\rho }^{2}\right) =\left(
929\,MeV\right) ^{2} 
\]
which is to be compared to the observed isoscalar vector mesons of $\omega
\left( 782\,MeV\right) $ and $\phi \left( 1020\,MeV\right) .$ This implies a
much more substantial mixing. The diagonalization of the corresponding mass
matrix: 
\[
\left( 
\begin{array}{cc}
m_{8}^{*2} & m_{08}^{*2} \\ 
m_{80}^{*2} & m_{0}^{*2}
\end{array}
\right) \longrightarrow \left( 
\begin{array}{cc}
m_{\omega }^{2} & 0 \\ 
0 & m_{\phi }^{2}
\end{array}
\right) 
\]
requires a mixing angle of 
\begin{equation}
\sin ^{2}\theta _{V}=\frac{m_{8}^{*2}-m_{\omega }^{2}}{m_{\phi
}^{2}-m_{\omega }^{2}}\;\;\;\;\;\;\;\;\;\;\text{or\ \ \ }\theta _{V}\simeq
50^{o}.  \label{vec-mix}
\end{equation}
The physical states should then be 
\begin{equation}
\left| \omega \right\rangle =\cos \theta _{V}\left| 8\right\rangle +\sin
\theta _{V}\left| 0\right\rangle \;\;\;\;\;\;\;\;\;\left| \phi \right\rangle
=-\sin \theta _{V}\left| 8\right\rangle +\cos \theta _{V}\left|
0\right\rangle .  \label{vec-mix-1}
\end{equation}
After substituting in Eqs.(\ref{8and0-combs}) and (\ref{vec-mix}) into Eq.(%
\ref{vec-mix-1}), we have 
\[
\left| \omega \right\rangle =0.7045\left| u\bar{u}+d\bar{d}\right\rangle
+0.0857\left| s\bar{s}\right\rangle \;\;\;\;\;\;\;\;\;\left| \phi
\right\rangle =-0.06\left| u\bar{u}+d\bar{d}\right\rangle +0.996\left| s\bar{%
s}\right\rangle . 
\]
This shows that $\omega $ has little $s$ quarks, while the $\phi $ mesons is
vector meson composed almost purely of $s$ quarks. Such a combination is
close to the situation of ``ideal mixing'', corresponding to an angle of $%
\theta _{0}\simeq 55^{o},$ with the non-strange and strange quarks being
completely separated: 
\begin{equation}
\left| \omega \right\rangle =\frac{1}{\sqrt{2}}\left| u\bar{u}+d\bar{d}%
\right\rangle \;\;\;\;\;\;\;\;\;\left| \phi \right\rangle =\left| s\bar{s}%
\right\rangle .
\end{equation}

\paragraph{The OZI rule}

It is observed experimentally that the $\phi $ meson decay predominantly
into strange-quark-bearing final states, even though the phase space, with $%
m_{\phi }>m_{\omega },$ favors its decay into nonstrange pions final states: 
\[
\begin{array}{lllllll}
\omega & \rightarrow 3\pi & \text{89\%} &  & \phi & \rightarrow K\bar{K} & 
\text{83\%} \\ 
&  &  &  &  & \rightarrow \rho \pi & \text{13\%} \\ 
&  &  &  &  & \rightarrow 3\pi & \ \text{3\%}
\end{array}
\]
with a ratio of partial decay widths $\Gamma (\phi \rightarrow 3\pi )/\Gamma
(\omega \rightarrow 3\pi )=0.014.\;$

This property of the hadron decays has been suggested to imply a strong
interaction regularity: \emph{the OZI-rule} --- the annihilations of the $s%
\bar{s}$ pair \emph{via} strong interaction are suppressed\cite{OZI}. We
remark that this suppression should be interpreted as a suppression of the
coupling strength rather than a phase space suppression due to the larger
strange quark mass ($i.e.$ it is above and beyond the conventional flavor
SU(3) breaking effect.)

The extension of the OZI-rule to heavy quarks of charm and bottom has been
highly successful. For example it explains the extreme narrowness of the
observed $J/\psi $ width because this $\left( c\bar{c}\right) $ bound state
is forbidden to decay into the OZI-allowed channel of $D\bar{D}\;$because,
with a mass of $m_{J/\psi }\simeq 3100\,MeV,$ it lies below the threshold of 
$2m_{D}\simeq 3700\,MeV.$

From the viewpoint of QCD, applications of the OZI-rule to the heavy $c,b,$
and $t$ quarks are much less controversial than those for strange quarks ---
even thought the rule was originally ``discovered'' in the processes
involving $s$ quarks. For heavy quarks, this can be understood in terms of
perturbative QCD and asymptotic freedom\cite{AP}. It is not the case for the 
$s$ quark which, as evidenced by the success of flavor-SU(3) symmetry,
should be considered a light quark. Furthermore, the phenomenological
applications of the OZI to strange quark processes have not been uniformly
successful. In contrast to the case of vector mesons Eq.(\ref{vec-mix}),
there is no corresponding success for the pseudoscalar mesons --- as
evidenced by the strong deviation from ideal mixing in the $\eta $ and $\eta
^{\prime }$ meson system, Eq.(\ref{ps-mix}).

\subsubsection{The OZI rule and the strange quark content of the nucleon}

A straightforward application of the $s$ quark OZI rule to the baryon is the
statement that operators that are bilinear in strange quark fields should
have a strongly suppressed matrix elements when taken between nonstrange
hadron states such as the nucleon. In particular we expect the fraction of $%
s $ quarks in a nucleon, Eq.(\ref{frac-def}), should be vanishingly small. 
\begin{equation}
F\left( s\right) =\frac{s+\bar{s}}{\sum \left( q+\bar{q}\right) }=\frac{%
\left\langle N\left| \bar{s}s\right| N\right\rangle }{\left\langle N\left| 
\bar{u}u+\bar{d}d+\bar{s}s\right| N\right\rangle }\simeq 0.  \label{sig0}
\end{equation}

The ``measured'' value of the pion-nucleon sigma term\cite{CDsigma}: 
\begin{equation}
\sigma _{\pi N}=m_{n}\left\langle N\left| \bar{u}u+\bar{d}d\right|
N\right\rangle  \label{sigma}
\end{equation}
and the SU(3) relation 
\begin{eqnarray}
M_{8} &\equiv &m_{8}\left\langle N\left| u_{8}\right| N\right\rangle =\frac{1%
}{3}\left( m_{n}-m_{s}\right) \left\langle N\left| \bar{u}u+\bar{d}d-2\bar{s}%
s\right| N\right\rangle  \nonumber \\
&=&M_{\Lambda }-M_{\Xi }\simeq -200\,MeV,  \label{M8}
\end{eqnarray}
which is obtained from Eqs.(\ref{u8}) and (\ref{bary-para}) in the isospin
invariant limit $\left( m_{3}u_{3}=0\right) ,$ allow us to make a
phenomenological estimate of the strange quark content of the nucleon\cite
{Cheng76}: We can rewrite the expression in Eq.(\ref{sig0}) as 
\begin{eqnarray}
F\left( s\right) &=&\frac{\left\langle N\left| \left( \bar{u}u+\bar{d}%
d\right) -\left( \bar{u}u+\bar{d}d-2\bar{s}s\right) \right| N\right\rangle }{%
\left\langle N\left| 3\left( \bar{u}u+\bar{d}d\right) -\left( \bar{u}u+\bar{d%
}d-2\bar{s}s\right) \right| N\right\rangle }  \nonumber \\
&=&\frac{\sigma _{\pi N}-25\,MeV}{3\sigma _{\pi N}-25\,MeV}  \label{cheng76}
\end{eqnarray}
where we have used (\ref{M8}) and the current quark mass ratio $%
m_{8}/m_{s}=-8$ corresponding to$\,m_{s}/m_{n}=25\;$of Eq.(\ref{n-s-ratio})$%
. $ Thus the validity of OZI rule, $F\left( s\right) =0$, would predict,
through (\ref{cheng76}), that $\sigma _{\pi N}$ should have a value close to 
$25\,MeV.$ However, the commonly accepted phenomenological value\cite
{sigma45} is more like $45\,MeV,$ which translates into a significant
strange quark content in the nucleon: 
\begin{equation}
F\left( s\right) \simeq 0.18.  \label{sfrac}
\end{equation}
We should however keep in mind that this number is deduced by using flavor $%
SU\left( 3\right) $ symmetry. Hence the kinematical suppression effect of $%
M_{s}>M_{n}$ has not been taken into account.

\section{Deep Inelastic Scatterings}

\subsection{Polarized lepton-nucleon scatterings}

There is a large body of work on the topic of probing the proton spin
structure through polarized deep inelastic scattering (DIS) of leptons on
nucleon target. The reader can learn more details by starting from two
excellent reviews of \cite{HYC-Lec} and \cite{Ma-Lec}.

\subsubsection{Kinematics and Bjorken scaling}

For a lepton (electron or muon) scattering off a nucleon target to produce
some hadronic final state $X,\;via$ the exchange of a photon (4-momentum $%
q_{\mu }$), the inclusive cross section can be written as a product 
\begin{equation}
d\sigma \left( l+N\rightarrow l+X\right) \propto l^{\mu \nu }W_{\mu \nu }
\end{equation}
where $l^{\mu \nu }$ is the known leptonic part while $W_{\mu \nu }$ is the
hadronic scattering amplitude squared, $\sum_{X}\left| T\left( \gamma
^{*}\left( q\right) +N\left( p\right) \rightarrow X\right) \right| ^{2},$
which is given, according to the optical theorem, by the imaginary part of
the forward Compton amplitude$:$%
\begin{eqnarray}
&&W_{\mu \nu }=\frac{1}{2\pi }\func{Im}\int \left\langle p,s\left| T\left(
J_{\mu }^{em}\left( x\right) J_{\nu }^{em}\left( 0\right) \right) \right|
p,s\right\rangle e^{iq\cdot x}d^{4}x  \nonumber \\
&=&\left( -g_{\mu \nu }+\frac{q_{\mu }q_{\nu }}{q^{2}}\right) F_{1}\left(
q^{2},\nu \right) +\left( p_{\mu }-\frac{p\cdot q}{q^{2}}q_{\mu }\right)
\left( p_{\nu }-\frac{p\cdot q}{q^{2}}q_{\nu }\right) \frac{F_{2}\left(
q^{2},\nu \right) }{p\cdot q}  \nonumber \\
&&+i\epsilon _{\mu \nu \alpha \beta }q^{\alpha }\left[ s^{\beta }\frac{%
g_{1}\left( q^{2},\nu \right) }{p\cdot q}+p\cdot q\,s^{\beta }-s\cdot
q\,p^{\beta }\frac{g_{2}\left( q^{2},\nu \right) }{\left( p\cdot q\right)
^{2}}\right]  \label{Wmunu}
\end{eqnarray}
where 
\begin{equation}
q^{2}\equiv -Q^{2}<0\;\;\;\;\text{and}\;\;\;\;\nu =\frac{p\cdot q}{M}
\label{q2nu}
\end{equation}
$M$ being the nucleon mass. $s^{\alpha }=\bar{u}_{N}\left( p,s\right) \gamma
^{\alpha }\gamma _{5}u_{N}\left( p,s\right) $ is the spin-vector of the
proton, and the variable $\nu $ is the energy loss of the lepton, $\nu
=E-E^{\prime }$. We have defined the spin-independent $F_{1,2}\left(
q^{2},\nu \right) $ and the spin-dependent $g_{1,2}\left( q^{2},\nu \right) $
\emph{structure functions}$.\;$In particular, the cross section asymmetry
with the target nucleon spin being anti-parallel and parallel to the beam of
longitudinally polarized leptons is given by the structure function $g_{1}:$ 
\begin{equation}
\frac{d\sigma ^{\uparrow \downarrow }}{dxdy}-\frac{d\sigma ^{\uparrow
\uparrow }}{dxdy}=\frac{e^{4}ME}{\pi Q^{4}}xy\left( 2-y\right) g_{1}+O\left( 
\frac{M^{2}}{Q^{2}}\right)  \label{spin-asym}
\end{equation}
where $x=\frac{Q^{2}}{2\nu M}$ and $y=\frac{\nu }{E}.$ In practice one
measures $g_{1}$ \emph{via} the (longitudinal) \emph{spin-asymmetry, } 
\begin{equation}
\mathcal{A}_{1}=\frac{d\sigma ^{\uparrow \uparrow }-d\sigma ^{\uparrow
\downarrow }}{d\sigma ^{\uparrow \uparrow }+d\sigma ^{\uparrow \downarrow }}%
\simeq 2x\frac{g_{1}}{F_{2}}.  \label{A1-asym}
\end{equation}
in the kinematic regime of $\nu \gg \sqrt{Q^{2}}.$

To probe the nucleon structure at small distance scale we need to go to the
large energy and momentum-transfer \emph{deep inelastic region} --- large \ $%
Q^{2}\;$and\ $\nu ,$ with fixed\ $x$. In the configuration space, this
corresponds to the lightcone regime. The statement of \emph{Bjorken scaling}
is that, in this kinematic limit, the structure functions approach
non-trivial functions of one variable: 
\begin{equation}
F_{1,2}\left( q^{2},\nu \right) \rightarrow F_{1,2}\left( x\right)
,\;\;\;\;\;\;\;g_{1,2}\left( q^{2},\nu \right) \rightarrow g_{1,2}\left(
x\right) .  \label{bj-scaling}
\end{equation}
Such problems can be studied with the formal approach of \emph{operator
product expansion}, which has a firm field theoretical-foundation in QCD, or
the more intuitive approach of \emph{parton model}, which can lead to
considerable insight about the hadronic structure.

\subsubsection{Inclusive sum rules via operator product expansion}

The forward Compton amplitude $T_{\mu \nu }$ is the matrix element, taken
between the nucleon states 
\begin{equation}
T_{\mu \nu }=\left\langle p,s\left| t_{\mu \nu }\right| p,s\right\rangle ,
\label{Tmunu}
\end{equation}
of the time-order product of two electromagnetic current operators 
\begin{equation}
t_{\mu \nu }=i\int d^{4}x\,e^{iq\cdot x}T\left( J_{\mu }\left( x\right)
J_{\nu }\left( 0\right) \right) .  \label{tmunu}
\end{equation}
It is useful to express the product of two operators at short distances as
an infinite series of local operators, $\mathcal{O}_{A}\left( x\right) 
\mathcal{O}_{B}\left( 0\right) =\sum_{i}C_{i}\left( x\right) \mathcal{O}%
_{i}\left( 0\right) $,\ as it is considerably simpler to work with the
matrix elements of local operators $\mathcal{O}_{i}\left( 0\right) $. For
DIS study we are interested in the lightcone limit $x^{2}\rightarrow 0.$
Hence operators of all possible dimensions $\left( d_{i}\right) $ and spins $%
\left( n\right) $ are to be included: 
\begin{equation}
\mathcal{O}_{A}\left( x\right) \mathcal{O}_{B}\left( 0\right)
=\sum_{i,n}C_{i}\left( x^{2}\right) x_{\mu _{1}}...x_{\mu _{n}}\mathcal{O}%
_{i}^{\mu _{1}...\mu _{n}}\left( 0\right)  \label{OPEspin-op}
\end{equation}
where $\mathcal{O}_{i}^{\mu _{1}...\mu _{n}}\left( 0\right) $ is understood
to be a symmetric traceless tensor operator (corresponding to a spin $n$
object). From dimension analysis we see that the coefficient 
\[
C_{i}\left( x^{2}\right) \sim \left( \sqrt{x^{2}}\right) ^{\tau
_{i}-d_{A}-d_{B}} 
\]
where $\tau _{i}=d_{i}-n$ is the \emph{twist} of the local operator $%
\mathcal{O}_{i}^{\mu _{1}...\mu _{n}}\left( 0\right) .\;$Thus in the
lightcone limit $x^{2}\rightarrow 0,$ the most important contributions come
from those operators with the lowest twist values.

In the short distance scale, the QCD running coupling is small so that
perturbation theory is applicable. In this way the c-numbers coefficients $%
C_{i}\left( x^{2}\right) $ can be calculated with the local operators $%
\mathcal{O}_{i}^{\mu _{1}...\mu _{n}}\left( 0\right) $ being the composite
operators of the quark and gluon fields.

We are interested, as in Eq.(\ref{tmunu}), in the operator products in the
momentum space. Namely, the above discussion has to be Fourier transformed
from configuration space into the momentum space: $x\rightarrow q,$ with the
relevant limit being $Q^{2}\rightarrow \infty .$ The spin-dependent case
corresponds to an operator product antisymmetric in the Lorentz indices $\mu 
$ and $\nu :$%
\begin{equation}
t_{\left[ \mu \nu \right] }=\sum_{\psi ,n=1,3,...}C_{\left( 3\right) }\left(
q^{2},\alpha _{s}\right) \left( \frac{2}{-q^{2}}\right) ^{n}i\epsilon _{\mu
\nu \alpha \beta }q^{\alpha }q_{\mu _{2}}....q_{\mu _{n}}\mathcal{O}_{A,\psi
}^{\beta \mu _{2}....\mu _{n}}  \label{OPE-t}
\end{equation}
where $C_{\left( 3\right) }\left( q^{2},\alpha _{s}\right) =1+O\left( \alpha
_{s}\right) ,$ [the subscript (3) reminds us of others terms, 1 \& 2, that
contribute to the spin-independent amplitudes $F_{1,2}$]. $\mathcal{O}%
_{A,\psi }^{\beta \mu _{2}....\mu _{n}}$ is a twist-two pseudotensor
operator: 
\begin{equation}
\mathcal{O}_{A,\psi }^{\beta \mu _{2}....\mu _{n}}=e_{\psi }^{2}\left( \frac{%
i}{2}\right) ^{n-1}\bar{\psi}\gamma ^{\beta }\stackrel{\longleftrightarrow }{%
D}^{\mu _{2}}....\stackrel{\longleftrightarrow }{D}^{\mu _{n}}\gamma _{5}\psi
\label{loc-op}
\end{equation}
where $\psi \;$is the quark field with charge $e_{\psi }$. The crossing
symmetry property 
\begin{equation}
t_{\mu \nu }\left( p,q\right) =t_{\nu \mu }\left( p,-q\right)
\label{cross-sym}
\end{equation}
implies that only odd-$n\,$ terms appear in the $\left[ \mu \nu \right] $
series. (By the same token, only even-$n$ terms contribute to the
spin-independent structure function $F_{1,2}.$)

The spin-dependent part of the forward Compton amplitude Eq.(\ref{Tmunu}) is 
\begin{equation}
T_{\left[ \mu \nu \right] }=\left\langle p,s\left| t_{\left[ \mu \nu \right]
}\right| p,s\right\rangle =i\epsilon _{\mu \nu \alpha \beta }q^{\alpha
}s^{\beta }\frac{\tilde{g}_{1}\left( q^{2},\nu \right) }{p\cdot q}+...
\label{T-g-tilda}
\end{equation}
Namely, $\func{Im}\tilde{g}_{1}\left( q^{2},\nu \right) =2\pi g_{1}\left(
q^{2},\nu \right) .\;$When we sandwich the OPE terms Eqs.(\ref{OPE-t}) and (%
\ref{loc-op}) into the nucleon states we need to evaluate matrix element 
\begin{equation}
\left\langle p,s\left| \mathcal{O}_{A,\psi }^{\beta \mu _{2}....\mu
_{n}}\right| p,s\right\rangle =2e_{\psi }^{2}A_{n,\psi }s^{\beta }p^{\mu
_{2}}....p^{\mu _{n}}  \label{An-def}
\end{equation}
Plug Eqs.(\ref{An-def}) and (\ref{OPE-t}) into Eq.(\ref{T-g-tilda}) we have 
\[
i\epsilon _{\mu \nu \alpha \beta }q^{\alpha }s^{\beta }\frac{\tilde{g}_{1}}{%
p\cdot q}=\sum_{n=1,3,...}^{\infty }C_{\left( 3\right) }\left( \frac{2}{%
-q^{2}}\right) ^{n}i\epsilon _{\mu \nu \alpha \beta }q^{\alpha }s^{\beta
}\left( p\cdot q\right) ^{n-1}2e_{\psi }^{2}A_{n,\psi }
\]
or 
\begin{equation}
\tilde{g}_{1}=\sum_{\psi ,n}2C_{\left( 3\right) }e_{\psi }^{2}A_{n,\psi
}\omega ^{n}  \label{OPEg1}
\end{equation}
where $\omega =\frac{2p\cdot q}{-q^{2}}$ is the inverse of the Bjorken-$x$
variable$.\;$Asymptotic freedom of QCD has allowed us to express the
structure function as a power series in $\omega ,$Eq.(\ref{OPEg1}) with
calculable c number coefficients $C_{\left( 3\right) }$ and ``unknown'' long
distance quantities $A_{n,\psi }.$ To turn this into a useful relation we
need to invert the summation over $n$ (\emph{i.e. }to\emph{\ }isolate the
coefficient $A_{n,\psi }$). For this we can use the Cauchy's theorem for
contour integration: 
\begin{equation}
\frac{1}{2\pi i}\oint d\omega \frac{\tilde{g}_{1}\left( \omega \right) }{%
\omega ^{n+1}}=\sum_{\psi }2C_{\left( 3\right) }e_{\psi }^{2}A_{n,\psi },
\label{cont-int}
\end{equation}
which can be related to physical processes by evaluating the LHS integral
with a deformed contour so that it wraps around the two physical cuts, $%
\omega =\left( 1,\infty \right) $ and $\left( -\infty ,-1\right) .$ (The
second region corresponding to the cross-channel process.) Using 
\begin{equation}
\tilde{g}_{1}\left( \omega +i\varepsilon \right) -\tilde{g}_{1}^{*}\left(
\omega +i\varepsilon \right) =2i\func{Im}\tilde{g}_{1}\left( \omega \right)
=4i\pi g_{1}\left( \omega \right) 
\end{equation}
and the crossing symmetry property 
\begin{equation}
g_{1}\left( p,q\right) =-g_{1}\left( p,-q\right) \text{ \ \ \ \ \ or\ \ \ \
\ }g_{1}\left( \omega ,q^{2}\right) =-g_{1}\left( -\omega ,q^{2}\right) ,
\label{g1cross}
\end{equation}
we then obtain 
\begin{eqnarray}
&&\frac{1}{2\pi i}\oint d\omega \frac{\tilde{g}_{1}\left( \omega \right) }{%
\omega ^{n+1}}=\frac{1}{\pi }\int_{1}^{\infty }d\omega \frac{\func{Im}\tilde{%
g}_{1}\left( \omega \right) }{\omega ^{n+1}}+\frac{1}{\pi }\int_{-\infty
}^{-1}d\omega \frac{\func{Im}\tilde{g}_{1}\left( \omega \right) }{\omega
^{n+1}}  \nonumber \\
&=&2\left[ 1-\left( -1\right) ^{n}\right] \int_{1}^{\infty }d\omega \frac{%
g_{1}\left( \omega \right) }{\omega ^{n+1}}=4\int_{0}^{1}x^{n-1}g_{1}\left(
x\right) dx.  \label{cont-int1}
\end{eqnarray}
We recall that the spin-index $n$ must be odd. The first-moment $\left(
n=1\right) $ sum 
\begin{equation}
\int_{0}^{1}dxg_{1}\left( x,Q^{2}\right) =\frac{1}{2}\sum_{\psi }C_{\left(
3\right) }e_{\psi }^{2}A_{1,\psi }  \label{g1sum}
\end{equation}
is of particular interest because the corresponding matrix element on the
RHS can be measured independently, \emph{Cf. }Eqs.(\ref{delq-def}) and (\ref
{An-def}): 
\begin{equation}
2A_{1}s^{\beta }=\left\langle p,s\left| \bar{\psi}\gamma ^{\beta }\gamma
_{5}\psi \right| p,s\right\rangle \equiv 2s^{\beta }\Delta \psi .
\label{delq-def1}
\end{equation}
Without including the higher order QCD corrections in the coefficient, we
have the\emph{\ g}$_{1}$\emph{\ sum rule} for the electron proton
scattering: 
\begin{equation}
\int_{0}^{1}dxg_{1}^{p}\left( x,Q^{2}\right) =\frac{1}{2}\left( \frac{4}{9}%
\Delta u+\frac{1}{9}\Delta d+\frac{1}{9}\Delta s\right) .  \label{g1sumrule}
\end{equation}
For the difference between scatterings on the proton and the neutron
targets, we can use the isospin relations $\left( \Delta u\right)
_{n}=\Delta d$ and $\left( \Delta d\right) _{n}=\Delta u$ to get: 
\begin{equation}
\int_{0}^{1}dx\left[ g_{1}^{p}\left( x,Q^{2}\right) -g_{1}^{n}\left(
x,Q^{2}\right) \right] =\frac{1}{6}\left( \Delta u-\Delta d\right) .
\label{Bjsum}
\end{equation}
The matrix element on the RHS: 
\begin{eqnarray}
2s^{\beta }\left( \Delta u-\Delta d\right)  &=&\left\langle p,s\left| \bar{u}%
\gamma ^{\beta }\gamma _{5}u-\bar{d}\gamma ^{\beta }\gamma _{5}d\right|
p,s\right\rangle   \nonumber \\
&=&\left\langle p,s\left| \bar{u}\gamma ^{\beta }\gamma _{5}d\right|
n,s\right\rangle =2s^{\beta }g_{A}  \label{gA}
\end{eqnarray}
is simply the axial vector decay constant of neutron beta decay. Including
the higher order QCD correction to the OPE Wilson coefficient, one can then
write down the \emph{Bjorken sum rule}: 
\begin{equation}
\int_{0}^{1}dx\left[ g_{1}^{p}\left( x,Q^{2}\right) -g_{1}^{n}\left(
x,Q^{2}\right) \right] =\frac{g_{A}}{6}C_{\left( NS\right) }  \label{Bjsum1}
\end{equation}
with the non-singlet coefficient\cite{qcd-ns-corr}, 
\begin{equation}
C_{\left( NS\right) }=1-\frac{\alpha _{s}}{\pi }-\frac{43}{12}\left( \frac{%
\alpha _{s}}{\pi }\right) ^{2}-20.22\left( \frac{\alpha _{s}}{\pi }\right)
^{3}+...  \label{QCDcor-ns}
\end{equation}

All experimental data are consistent with this theoretical prediction.

\begin{description}
\item[\textbf{Remark\ }]  \textbf{\ }\emph{Anomalous dimension and the }$%
Q^{2}$\emph{-dependence: }The $Q^{2}$-dependence of the moment integral,
such as LHS of Eq.(\ref{g1sum}), are given by $\alpha _{s}\left(
Q^{2}\right) \sim 1/\ln Q^{2}$ in the coefficient function and by the $Q^{2}$%
-evolution of the operator according to the renormalization group equation%
\cite{CL-10}, which yields 
\begin{equation}
\frac{\left\langle p,s\left| \left. \mathcal{O}\right| _{Q}\right|
p,s\right\rangle }{\left\langle p,s\left| \left. \mathcal{O}\right|
_{Q_{0}}\right| p,s\right\rangle }=\left[ \frac{\alpha _{s}\left( Q\right) }{%
\alpha _{s}\left( Q_{0}\right) }\right] ^{\frac{\gamma }{2b}}
\label{ano-dim}
\end{equation}
where $\gamma $ is the anomalous dimension of the operator $\mathcal{O}$ and 
$b$ is the leading coefficient in the QCD $\beta $ function. The label $Q$
in the matrix elements refers to the mass scale at which the operator is
renormalized, chosen at $\mu ^{2}\simeq Q^{2}$ in order to avoid large
logarithms. For the $g_{1}$ sum rule Eq.(\ref{g1sum}) the $Q^{2}$-dependence
is particularly simple. The non-singlet axial current is (partially)
conserved, hence has anomalous dimension $\gamma =0.$ The singlet current is
not conserved because of axial anomaly (see discussion below).\ But it has
very weak $Q^{2}$-dependence because the corresponding anomalous dimension
starts at the two-loop level.
\end{description}

\subsubsection{The parton model approach}

The $g_{1}$ sum rule of Eq.(\ref{g1sumrule}) has been derived directly
through OPE from QCD. We can also get this result by using the parton model,
which pictures the target hadron, in the infinite momentum frame, as
superposition of quark and gluon partons each carrying a fraction $\left(
x\right) \;$of the hadron momentum. For the short distance processes one can
calculate the reaction cross section as an incoherent sum over the rates for
the elementary processes. Thus in Compton scattering, a photon (momentum $%
q_{\mu }$) strikes a parton ($xp_{\mu }$) turning it into a final state
parton ($q_{\mu }+xp_{\mu }$), the initial and final partons must be on
shell: 
\begin{equation}
\left( xp_{\mu }\right) ^{2}=\left( q_{\mu }+xp_{\mu }\right) ^{2}\;\;\;\;\;%
\text{or}\;\;\;\;\;x=\frac{-q^{2}}{2p\cdot q}.  \label{parton-x}
\end{equation}
Hence the Bjorken-$x$ variable has the interpretation as the fraction of the
longitudinal momentum carried by the parton. A simple calculation\cite{CL-7}
shows the scaling structure functions being directly related to the density
of partons with momentum fraction $x:$%
\begin{equation}
F_{2}^{p}\left( x\right) =x\sum_{q=u,d,s}e_{q}^{2}\left[ q\left( x\right) +%
\bar{q}\left( x\right) \right]  \label{parton-F2}
\end{equation}
and 
\begin{eqnarray}
g_{1}^{p}\left( x\right) &=&\frac{1}{2}\sum_{q=u,d,s}e_{q}^{2}\left[
q_{+}\left( x\right) -q_{-}\left( x\right) +\bar{q}_{+}\left( x\right) -\bar{%
q}_{-}\left( x\right) \right]  \nonumber \\
&=&\frac{1}{2}\sum e_{q}^{2}\left[ \Delta _{q}\left( x\right) +\Delta _{\bar{%
q}}\left( x\right) \right] =\frac{1}{2}\sum e_{q}^{2}\Delta q\left( x\right)
\label{parton-g1}
\end{eqnarray}
Thus the spin asymmetry of Eq.(\ref{A1-asym}) has the interpretation as 
\begin{equation}
\mathcal{A}_{1}\left( x\right) \simeq \frac{\sum_{q}e_{q}^{2}\left[ \Delta
_{q}\left( x\right) +\Delta _{\bar{q}}\left( x\right) \right] }{%
\sum_{q}e_{q}^{2}\left[ q\left( x\right) +\bar{q}\left( x\right) \right] }.
\label{part-A1}
\end{equation}

Comparing this interpretation of the spin-dependent structure function to
that for the proton matrix elements of the axial vector current Eq.(\ref
{delq-def}), we see that the $g_{1}$ sum rule Eq.(\ref{g1sumrule}) implies
the consistency condition of 
\begin{equation}
\int_{0}^{1}q_{\pm }\left( x\right) dx=q_{\pm }\;\;\;\;\;\;\;\int_{0}^{1}%
\bar{q}_{\pm }\left( x\right) dx=\bar{q}_{\pm }.  \label{par-g1sim}
\end{equation}
In other words, the proton matrix element of the local axial vector current $%
\left\langle p,s\left| \mathcal{O}_{A,q}\right| p,s\right\rangle $ can be
evaluated, in the partonic language, by taking the axial vector current
between quark states ($\left\langle q,h\left| \mathcal{O}_{A,q}\right|
q,h\right\rangle =2h$) and multiplying it by the probability of finding the
quark in the target proton: 
\begin{equation}
\left\langle p,s\left| \mathcal{O}_{A,q}\right| p,s\right\rangle
=\sum_{q,h}\left\langle q,h\left| \mathcal{O}_{A,q}\right| q,h\right\rangle
q_{h}(x)=\left( \Delta q\right) _{p}  \label{q-con-axial}
\end{equation}
where $\left( \Delta q\right) _{p}\equiv \Delta q$%
\begin{equation}
\Delta q\left( x\right) =q_{+}\left( x\right) -q_{-}\left( x\right) +\bar{q}%
_{+}\left( x\right) -\bar{q}_{-}\left( x\right) \equiv \Delta _{q}\left(
x\right) +\Delta _{\bar{q}}\left( x\right) .  \label{q-spin}
\end{equation}

\paragraph{Ellis-Jaffe sum rule and the phenomenological values of $\Delta q$%
}

Besides 
\begin{equation}
\Delta u-\Delta d=g_{A}=F+D=1.2573\pm 0.0028,  \label{del3}
\end{equation}
if we assume flavor SU(3) symmetry, we can fix another octet combination 
\begin{equation}
\Delta u+\Delta d-2\Delta s=\Delta _{8}=3F-D=0.601\pm 0.038  \label{del8}
\end{equation}
which can be gotten by fitting the axial vector couplings of the hyperon
beta decays\cite{axialFD}. In this way Eq.(\ref{g1sumrule}) can be written
as 
\begin{equation}
\Gamma _{p}=\int_{0}^{1}dxg_{1}^{p}\left( x\right) =\frac{C_{\left(
NS\right) }}{36}\left( 3g_{A}+\Delta _{8}\right) +\frac{C_{\left( S\right) }%
}{9}\Delta \Sigma  \label{g1sumrule1}
\end{equation}
where $\Delta \Sigma =\Delta u+\Delta d+\Delta s.$ The non-singlet
coefficient has been displayed in Eq.(\ref{QCDcor-ns}) while the singlet
term has been calculated to be\cite{qcd-s-corr} 
\begin{equation}
C_{\left( S\right) }=1-\frac{\alpha _{s}}{\pi }-1.0959\left( \frac{\alpha
_{s}}{\pi }\right) ^{2}+...  \label{QCDcor-s}
\end{equation}

If one assume $\Delta s=0,\;$thus $\Delta \Sigma =\Delta _{8\;}$we then
obtain the \emph{Ellis-Jaffe sum rule}\cite{ejsum}\emph{\ }with the RHS of
Eq.(\ref{g1sumrule1}) expected (for $\alpha _{s}\simeq 0.25$) to be around $%
0.175,$ had become the baseline of expectation for the spin-dependent DIS.
The announcement by EMC collaboration in the late 1980's that it had
extended the old SLAC result\cite{oldSLAC} to new kinematic region and
obtained an experimental value for $\Gamma _{p}$ deviated significantly from
the Ellis-Jaffe value\cite{emc} had stimulated a great deal of activity in
this area of research. In particular another generation of polarized DIS on
proton and neutron targets have been performed by SMC at CERN\cite{smc} and
by E142-3 at SLAC\cite{e142-3}. The new data supported the original EMC
findings of $\Delta s\neq 0$ and a much-less-than-unity of the total spin
contribution $\Delta \Sigma \ll 1,$ although the magnitude was not as small
as first thought. The present experimental result may be summarized as\cite
{ellisk} 
\begin{eqnarray}
\Delta u &=&\;0.82\pm 0.06,\;\;\;\Delta d=-0.44\pm 0.06,  \label{delq-expt}
\\
\Delta s &=&-0.11\pm 0.06,\;\;\;\Delta \Sigma =\;0.27\pm 0.11.  \nonumber
\end{eqnarray}

The deviation from the simple quark model prediction Eq.(\ref{sqm-delq}) 
\begin{equation}
\left( \Delta q\right) _{\text{exptl}}<\left( \Delta q\right) _{\text{sQM}}
\label{qseapol}
\end{equation}
indicates a quark sea strongly polarized in the opposite direction from the
proton spin. That the total quark contribution is small means that the
proton spin is built up from other components such as orbital motion of the
quarks and, if in the relevant region, gluons.

\subsubsection{Axial vector current and the axial anomaly}

The most widely discussed interpretation of the proton spin problem is the
suggestion that the gluon may provide significant contribution \emph{via}
the axial anomaly\cite{Anomaly}. Let us first review some elementary aspects
of anomaly. The $SU\left( 3\right) _{color}$ gauge symmetry of QCD is of
course anomaly-free. The anomaly under discussion is the one associated with
the global axial $U\left( 1\right) $ symmetry. Namely, the $SU\left(
3\right) $-singlet axial current$\;A_{\mu }^{\left( 0\right) }=\sum_{q=u,d,s}%
\bar{q}\gamma _{\mu }\gamma _{5}q\;$has an anomalous divergence 
\begin{equation}
\partial ^{\mu }A_{\mu }^{\left( 0\right) }=\sum_{q=u,d,s}2m_{q}\left( \bar{q%
}i\gamma _{5}q\right) +n_{f}\frac{\alpha _{s}}{2\pi }trG^{\mu \nu }%
\widetilde{G}_{\mu \nu }  \label{ano-div}
\end{equation}
where $G^{\mu \nu }$ is the gluon field tensor, $\widetilde{G}_{\mu \nu }$
its dual. $n_{f}=3$ is the number of excited flavors. For our purpose it is
more convenient to express this in terms of each flavor separately. 
\begin{equation}
\partial ^{\mu }\left( \bar{q}\gamma _{\mu }\gamma _{5}q\right)
=2m_{q}\left( \bar{q}i\gamma _{5}q\right) +\frac{\alpha _{s}}{2\pi }trG^{\mu
\nu }\widetilde{G}_{\mu \nu }  \label{ano-div-q}
\end{equation}

Axial anomaly enters into the discussion of partonic contributions to the
proton spin as follows: Because anomaly, being related to the UV
regularization of the triangle diagram, is a short-distance phenomena, it
makes a hard, thus perturbatively calculable (though not the amount),
contribution from the gluon so that Eq.(\ref{q-con-axial}) is modified: 
\begin{equation}
\left\langle p,s\left| \mathcal{O}_{A,q}\right| p,s\right\rangle
=\sum_{q,h}\left\langle q,h\left| \mathcal{O}_{A,q}\right| q,h\right\rangle
Q_{h}(x)+\sum_{G,h}\left\langle G,h\left| \mathcal{O}_{A,q}\right|
G,h\right\rangle G_{h}\left( x\right)   \label{qg-con-axial}
\end{equation}
where $G_{h}$, just as the quark density $Q_{h}$ being given by Eq.(\ref
{q-spin}), is the spin-dependent gluonic density. The gluonic matrix element
of the axial vector current $\left\langle G,h\left| \mathcal{O}_{A,q}\right|
G,h\right\rangle $ is just the anomaly triangle diagram which, with $%
\left\langle q,h\left| \mathcal{O}_{A,q}\right| q,h\right\rangle $
normalized to $\pm 1$, yields a coefficient of $\mp \frac{\alpha _{s}}{2\pi }%
.$ In this way the proton matrix element of the axial vector current is
interpreted as being a sum of ``true'' quark spin contribution $\Delta Q$
and the gluon spin contribution:
\begin{equation}
\Delta q\left( x\right) =\Delta Q\left( x\right) -\frac{\alpha _{s}}{2\pi }%
\Delta G\left( x\right) ,  \label{q-glue-pspin}
\end{equation}
where $\Delta G\left( x\right) =G_{+}\left( x\right) -G_{-}\left( x\right)
.\;$Superficially, the second term is of higher order. But because the $\ln
Q^{2}\;$growth of $\Delta G\;$(due to gluon blemsstrahlung by quarks)
compensates for the running coupling $\alpha _{s}\sim \left( \ln
Q^{2}\right) ^{-1},$ the combination $\alpha _{s}\Delta G$ is independent of 
$Q^{2}$ at the leading order, and the gluonic contribution to the proton
spin may not be negligible. However in order to obtain the simple quark
model result of $\Delta S=0,$ a very large $\Delta G$ is required: 
\begin{equation}
-\frac{\alpha _{s}}{2\pi }\Delta G=\Delta s\simeq -0.1\;\;\Rightarrow
\;\;\Delta G\simeq 2.5.
\end{equation}

\subsubsection{Semi-inclusive polarized DIS}

From the \emph{inclusive} lepton nucleon scattering we are able to extract
the quark contribution to the proton spin, $\Delta q=\Delta _{q}+\Delta _{%
\bar{q}}.\;$Namely, we can only get the sum of the quark and antiquark
contributions together. More detailed information of the spin structure can
be obtained from polarized \emph{semi-inclusive} DIS, where in addition to
the scattered lepton some specific hadron $h$ is also detected. 
\[
l+N\rightarrow l+h+X 
\]
The (longitudinal) spin asymmetry of the inclusive process can be expressed
in terms of quark distributions as in Eq.(\ref{part-A1}): 
\begin{equation}
\mathcal{A}_{1}\simeq \frac{\sum_{q}e_{q}^{2}\left( \Delta _{q}+\Delta _{%
\bar{q}}\right) }{\sum_{q}e_{q}^{2}\left( q+\bar{q}\right) }
\label{partonA1}
\end{equation}
Similarly one can measure the spin-asymmetry measured in semi-inclusive
case: 
\begin{equation}
\mathcal{A}_{1}^{h}\simeq \frac{\sum_{q}e_{q}^{2}\left( \Delta
_{q}D_{q}^{h}+\Delta _{\bar{q}}D_{\bar{q}}^{h}\right) }{\sum_{q}e_{q}^{2}%
\left( qD_{q}^{h}+\bar{q}D_{\bar{q}}^{h}\right) }  \label{partonA1h}
\end{equation}
where $D_{q}^{h}$, the fragmentation function for a quark $q$ to produce the
hadron $h,$ is assumed to be spin-independent. Separating $\Delta _{\bar{q}}$
from $\Delta _{q}$ is possible because $D_{\bar{q}}^{h}\neq D_{q}^{h}$. For
example, given the quark contents such as $\pi ^{+}\sim \left( u\bar{d}%
\right) $ and $\pi ^{-}\sim \left( \bar{u}d\right) $, we expect 
\[
D_{u}^{\pi ^{+}}\gg D_{\bar{u}}^{\pi ^{+}},\;\;\;D_{\bar{d}}^{\pi ^{+}}\gg
D_{d}^{\pi ^{+}},\;\;\;\;\text{and\ \ \ \ }D_{u}^{\pi ^{-}}\ll D_{\bar{u}%
}^{\pi ^{-}},\;\;\;D_{\bar{d}}^{\pi ^{-}}\ll D_{d}^{\pi ^{-}} 
\]
In this way the SMC collaboration\cite{smc-semi} made a fit of their
semi-inclusive data, in the approximation of $\Delta _{\bar{u}}=\Delta _{%
\bar{d}}$ and $\Delta _{s}=\Delta _{\bar{s}}\propto s\left( x\right) $ (the
strange quark distribution did not play an important role, and the final
result is insensitive to variation of $\Delta s$)$.\;$SMC was able to
conclude that the polarization of the non-strange antiquarks is compatible
with zero over the full range of $x:$%
\begin{equation}
\Delta _{\bar{u}}=\Delta _{\bar{d}}=-0.02\pm 0.09\pm 0.03
\label{smc-semi-result}
\end{equation}
This is to be compared to their result for $\widetilde{\Delta q}=\Delta
_{q}-\Delta _{\bar{q}}:$%
\[
\widetilde{\Delta u}=1.01\pm 0.19\pm 0.14\;\;\;\;\;\;\widetilde{\Delta d}%
=-0.57\pm 0.22\pm 0.11. 
\]
Namely, while the data from inclusive processes suggest that the quark sea
is strongly polarized --- as indicated by the large deviation of measured $%
\Delta q$ from their simple quark model prediction Eqs.(\ref{delq-expt}) and
(\ref{sqm-delq}), the SMC study of the semi-inclusive processes hints that
the antiquarks in the sea are not strongly polarized.

\subsubsection{Baryon magnetic moments}

One of the puzzling aspects of the proton spin problem is that, given the
significant deviation of the quark spin factors $\Delta q$ in Eq.(\ref
{delq-expt}) from the sQM values, it is hard to see how could the same $%
\left( \Delta q\right) _{sQM}$ values manage to yield such a good
description of the baryon magnetic moments, as shown in Table 1.

For this we can only give a partially satisfactory answer : If we assume
that the anitquarks in the proton sea is not polarized $\Delta _{\bar{q}}=0$%
, for which the SMC result Eq.(\ref{smc-semi-result}) gives some evidence
(and it is also a prediction of the chiral quark model to be discussed in
Sec. 3), we can directly use the $\Delta q$ of Eq.(\ref{delq-expt}) to
evaluate the polarization difference: $\widetilde{\Delta q}=\Delta
_{q}=\Delta q$ in Eq.(\ref{deldiff}).$\;$We can then attempt a fit of the
baryon magnetic moments in exactly the same way we had fit them by using $%
\left( \widetilde{\Delta q}\right) _{sQM}$ as in Table 1. The resultant fit,
surprisingly, is equally good --- in fact \emph{better,} in the sense of
lower $\chi ^{2}$\cite{Karl}\cite{CL96}. Namely, both the sQM $\Delta q$ and
experimental values of $\Delta q$ can, rather miraculously, fit the same
magnetic moment data. In this sense, the new spin structure poses no
intrinsic contradiction with respect to the magnetic moment phenomenology.

That it is possible to fit the same baryon magnetic moments with $\left( 
\widetilde{\Delta q}\right) _{sQM}$ and $\left( \widetilde{\Delta q}\right)
_{\text{exptl}}$ is due to the fact that the baryon moment, such as Eq.(\ref
{proton}), is a sum of products $\mu _{B}=\sum \widetilde{\Delta q}\,\mu _{q}
$ hence different $\left( \widetilde{\Delta q}\right) ^{\prime }$s can yield
the same $\mu _{B}$ if $\left( \mu _{q}\right) ^{\prime }$s are changed
correspondingly. In both cases we have $\mu _{u}=-2\mu _{d}$ and $\mu
_{s}\simeq -\frac{2}{3}\mu _{d}$. For the sQM case, we find $\mu _{d}\simeq
-0.9\,\mu _{N}$ while for the experimental $\Delta q$ case, we need $\mu
_{d}\simeq -1.4\,\mu _{N}.\;$This shift means a $35\%$ change in the
constituent quark mass value --- thus a $35\%$ difference with the
constituent quark mass value obtained from the baryon mass fit in Eq.(\ref
{cq-mass-ss}). Consequently, we regard the magnetic moment problem still as
an unsolved puzzle.

\subsection{DIS on proton vs neutron targets}

\subsubsection{Lepton-nucleon scatterings}

The spin-averaged nucleon structure function $F_{2}$ can be expressed in
terms of the quark densities as in Eq.(\ref{parton-F2}) 
\begin{eqnarray*}
F_{2}^{p}\left( x\right) &=&x\left[ \frac{4}{9}\left( u+\bar{u}\right) +%
\frac{1}{9}\left( d+\bar{d}\right) +\frac{1}{9}\left( s+\bar{s}\right)
\right] \\
&& \\
F_{2}^{n}\left( x\right) &=&x\left[ \frac{4}{9}\left( d+\bar{d}\right) +%
\frac{1}{9}\left( u+\bar{u}\right) +\frac{1}{9}\left( s+\bar{s}\right)
\right] ,
\end{eqnarray*}
where we have used the isospin relations of $\left( u\right) _{p}=\left(
d\right) _{n}$ and $\left( d\right) _{p}=\left( u\right) _{n}.$ Their
difference is 
\begin{eqnarray*}
\frac{1}{x}\left[ F_{2}^{p}\left( x\right) -F_{2}^{n}\left( x\right) \right]
&=&\frac{1}{3}\left[ \left( u-d\right) +\left( \bar{u}-\bar{d}\right) \right]
\\
&=&\frac{1}{3}\left[ 2\mathcal{I}_{3}+2\left( \bar{u}-\bar{d}\right) \right]
\end{eqnarray*}
where $\mathcal{I}_{3}=\frac{1}{2}\left[ \left( u-d\right) -\left( \bar{u}-%
\bar{d}\right) \right] $ with it integral being the third component of the
isospin: $\int_{0}^{1}dx\mathcal{I}_{3}\left( x\right) =\frac{1}{2}.\;$The
simple assumption that $\bar{u}=\bar{d}$ in the quark sea, which is
consistent with it being created by the flavor-independent gluon emission,
then leads the Gottfried sum rule\cite{gsr} 
\begin{equation}
I_{G}=\int_{0}^{1}\frac{dx}{x}\left[ F_{2}^{p}\left( x\right)
-F_{2}^{n}\left( x\right) \right] =\frac{1}{3}.  \label{Gsumrule}
\end{equation}
Experimentally, NMC found that, with a reasonable extrapolation in the very
small-$x$ region, the integral $I_{G}$ deviated significantly from one-third%
\cite{nmc}: 
\begin{equation}
I_{G}=0.235\pm 0.026=\frac{1}{3}+\frac{2}{3}\int_{0}^{1}\left[ \bar{u}\left(
x\right) -\bar{d}\left( x\right) \right] dx.  \label{NMCvalue}
\end{equation}
This translates into the statement that, in the proton quark sea, there are
more $d$-quark pairs as compared to the $u$-quark pairs. 
\begin{equation}
\bar{u}-\bar{d}=-0.147\pm 0.026.  \label{ubar-dbar}
\end{equation}

\begin{description}
\item[\textbf{Remark }]  \emph{Gottfried sum rule does not follow directly
from QCD without additional assumption.} Unlike the $g_{1}$ sum rule, the
Gottfried sum rule can not be derived from QCD \emph{via} \emph{operator
product expansion}. A simple way to see this: Because the spin-independent
structure function $F_{2}$ has opposite crossing symmetry property from that
of $g_{1},$ only even-$n$ terms can contribute. Hence there is no way to
obtain a non-trivial relation for the odd-$n$ moment sums of $F_{2}\;$(which
the Gottfried sum rule would be an example). But in the context of \emph{%
parton model}, the Gottfried sum provides us with an important measure of
the flavor structure of the proton quark sea.
\end{description}

\subsubsection{Drell-Yan processes}

Because to conclude that NMC data showing a violation of the Gottfried sum
rule one needs to make an extrapolation into the small-$x$ regime, an
independent confirmation of $\bar{u}\neq \bar{d}$ would be helpful. A
measurement of the difference of the Drell-Yan process of proton $%
pN\rightarrow l^{+}l^{-}X\;$on proton and neutron targets can detect the
antiquark density because in such a process the massive $\left(
l^{+}l^{-}\right) $ pair is produced by $\left( q\bar{q}\right) $
annihilations\cite{es-DY}.

Let us denote the differential cross sections as 
\begin{eqnarray}
\sigma ^{pN} &\equiv &\frac{d^{2}\sigma \left( pN\rightarrow
l^{+}l^{-}X\right) }{d\sqrt{\tau }dy}  \nonumber \\
&=&\frac{8\pi \alpha }{9\sqrt{\tau }}\sum_{q=u,d,s}e_{q}^{2}\left[
q^{P}\left( x_{1}\right) \bar{q}^{T}\left( x_{2}\right) +\bar{q}^{P}\left(
x_{1}\right) q^{T}\left( x_{2}\right) \right]  \label{DYdensity}
\end{eqnarray}
where $\sqrt{\tau }=\frac{M}{\sqrt{s}}\;$with $\sqrt{s}$ being the CM
collision energy and $M$ is the invariant mass of the lepton pair. $y$ being
the rapidity, the fraction of momentum carried by the parton in the
projectile $\left( P\right) $ is given by $x_{1}=\sqrt{\tau }e^{y}$ and the
fraction in the target $\left( T\right) $ given by $x_{2}=\sqrt{\tau }%
e^{-y}.\;$Explicitly writing out the quark densities of Eq.(\ref{DYdensity}%
): 
\begin{eqnarray*}
\sigma ^{pp} &=&\frac{8\pi \alpha }{9\sqrt{\tau }}\left\{ \frac{4}{9}\left[
u\left( x_{1}\right) \bar{u}\left( x_{2}\right) +\bar{u}\left( x_{1}\right)
u\left( x_{2}\right) \right] +\frac{1}{9}\left[ d\left( x_{1}\right) \bar{d}%
\left( x_{2}\right) +\bar{d}\left( x_{1}\right) d\left( x_{2}\right) \right]
+s\text{ term}\right\} \\
&& \\
\sigma ^{pn} &=&\frac{8\pi \alpha }{9\sqrt{\tau }}\left\{ \frac{4}{9}\left[
u\left( x_{1}\right) \bar{d}\left( x_{2}\right) +\bar{u}\left( x_{1}\right)
d\left( x_{2}\right) \right] +\frac{1}{9}\left[ d\left( x_{1}\right) \bar{u}%
\left( x_{2}\right) +\bar{d}\left( x_{1}\right) u\left( x_{2}\right) \right]
+s\text{ term}\right\}
\end{eqnarray*}
In this way the \emph{DY cross section asymmetry} can be found: 
\begin{eqnarray}
A_{DY} &=&\frac{\sigma ^{pp}-\sigma ^{pn}}{\sigma ^{pp}+\sigma ^{pn}} 
\nonumber \\
&=&\frac{\left[ 4u\left( x_{1}\right) -d\left( x_{1}\right) \right] \left[ 
\bar{u}\left( x_{2}\right) -\bar{d}\left( x_{2}\right) \right] +\left[
u\left( x_{2}\right) -d\left( x_{2}\right) \right] \left[ 4\bar{u}\left(
x_{1}\right) -\bar{d}\left( x_{1}\right) \right] }{\left[ 4u\left(
x_{1}\right) +d\left( x_{1}\right) \right] \left[ \bar{u}\left( x_{2}\right)
+\bar{d}\left( x_{2}\right) \right] +\left[ u\left( x_{2}\right) +d\left(
x_{2}\right) \right] \left[ 4\bar{u}\left( x_{1}\right) +\bar{d}\left(
x_{1}\right) \right] }  \nonumber \\
&=&\frac{\left( 4\lambda -1\right) \left( \bar{\lambda}-1\right) +\left(
\lambda -1\right) \left( 4\bar{\lambda}-1\right) }{\left( 4\lambda +1\right)
\left( \bar{\lambda}+1\right) +\left( \lambda +1\right) \left( 4\bar{\lambda}%
+1\right) }
\end{eqnarray}
where $\lambda \left( x\right) =u\left( x\right) /d\left( x\right) $ and $%
\bar{\lambda}\left( x\right) =\bar{u}\left( x\right) /\bar{d}\left( x\right)
.\,\;$Thus with measurements of $A_{DY}$ and data fit for $\lambda $ in the
range of $\left( 2.0,2.7\right) ,\;$the NA51 Collaboration\cite{na51}
obtained, at kinematic point of $y=0$ and $x_{1}=x_{2}=x=0.18,$ the ratio of
antiquark distributions to be 
\begin{equation}
\bar{u}/\bar{d}=0.51\pm 0.04\pm 0.05  \label{na51result}
\end{equation}
confirming that there are more (by a factor of 2) \emph{d}-quark pairs than 
\emph{u}-quark pairs.

\section{The Proton Spin-Flavor Structure in the Chiral Quark Model}

\subsection{The naive quark sea}

A significant part of the nucleon structure study involves non-perturbative
QCD. As the structure problem may be very complicated when viewed directly
in terms of the fundamental degrees of freedom (current quarks and gluons),
it may well be useful to separate the problem into two stages. One first
identifies the relevant degrees of freedom (DOF) in terms of which the
description for such non-perturbative physics will be simple, intuitive and
phenomenologically correct; at the next stage, one then elucidates the
relations between these non-perturbative DOFs in terms of the QCD quarks and
gluons. Long before the advent of the modern gauge theory of strong
interaction, we have already gained insight into the nucleon structure with
the simple nonrelativistic constituent quark model (sQM). This model
pictures a nucleon as being a compound of three almost free \emph{u-} and 
\emph{d}-\emph{constituent quarks} (with masses, much larger than those of
current quarks, around a third of the nucleon mass) enclosed within some
simple confining potential. There are many supporting evidence for this
picture. We have reviewed some of this in Sec. 1. Also, the nucleon \
structure functions in the large momentum fraction \emph{x} region, where
the valence quarks are expected to be the dominant physical entities, are
invariably found to be compatible with them being evolved from a low $Q^{2}$
regime described by sQM. For this aspect of the quark model we refer the
reader to Ref.\cite{Close}.

However in a number of instances where small \emph{x} region can contribute
one finds the observed phenomena to be significantly different from these
sQM expectations. This has led many people to call sQM the ''\emph{naive}
quark model'' and to suggest a rethinking of the nucleon structure. But we
would argue that the approach is correct, and only the generally expected
features of the \emph{quark-sea }are too simple. This ''naive quark-sea''
(nQS) is supposed to be composed exclusively of the \emph{u} and \emph{d}
quark pairs. Namely, based on the notion of OZI rule, one would anticipate a
negligibly small presence of the strange quark pairs inside the nucleon.
This implies, as given in Eq.(\ref{cheng76}), a pion-nucleon sigma term
value of $\sigma _{\pi N}\simeq 25\,MeV.$ Furthermore, the similarity of the 
\emph{u} and \emph{d} quark masses and the flavor-independent nature of the
gluon couplings led some people to expect that \ $\overline{d}=\overline{u}$%
, thus to the validity of the Gottfried sum rule, Eq.(\ref{Gsumrule}).

In the sQM, there is no quark-sea and the proton spin is build up entirely
by the valence quark spins. We have deduced the quark contributions to the
proton spin as in Eq.(\ref{sqm-delq}), which leads to an axial-vector
coupling strength of $g_{A}=\Delta u-\Delta d=5/3.$\ If one introduces a
quark-sea, the nQS feature of $\overline{s}\simeq 0$ (thus $\Delta s\simeq 0$%
) leads us to the Ellis-Jaffe sum rule, $\int_{0}^{1}dxg_{1}^{p}\left(
x\right) =0.175.$

Phenomenologically none of these nQS features 
\[
\begin{tabular}{cc}
& \underline{Features of the naive quark sea} \\ 
$flavor:$ & $\;\text{ }\overline{s}=0\;\;\;\;\text{and \ \ \ \ \ }\overline{d%
}=\overline{u}$ \\ 
$spin:$ & $\;\;\;\;\Delta s=0\;\;\;\;\;\;\;\left( \Delta _{q}=\Delta _{\bar{q%
}}\right) _{sea}$%
\end{tabular}
\]
have been found to be in agreement with experimental observations. As far
back as 1976, the connection of the $\sigma _{\pi N}$ value to the strange
quark content of the nucleon has been noted. It was pointed out that the
then generally accepted phenomenological value of $60\,MeV$ differed widely
from the OZI expectation\cite{Cheng76}. In recent years, the $\sigma _{\pi N}
$ value has finally settled down to a more moderate value of $\sigma _{\pi
N}\simeq 45\,MeV$ when a more reliable calculation confirmed the existence
of a significant correction due to the two-pion cut\cite{sigma45}.
Nevertheless, this reduced value still translates into a nucleon strange
quark fraction of $0.18,$ see Eq.(\ref{sfrac}).

As for the proton spin, starting with EMC in the 1980's, the polarized DIS
experiments of leptons on proton target have shown that Ellis-Jaffe sum rule
is violated. The first moment the spin-dependent structure function $g_{1}$
has allowed us to obtain the individual $\Delta q\;$of Eq.(\ref{delq-expt}).
We have already noted that they are all less than the sQM values of Eq.(\ref
{sqm-delq}), suggesting that for each flavor the quark-sea is polarized
strongly in the opposite direction to the proton spin. 
\begin{eqnarray*}
\Delta q &=&\left( \Delta q\right) _{sQM}+\left( \Delta q\right)
_{sea}<\left( \Delta q\right) _{sQM} \\
&\Rightarrow &\;\;\;\left( \Delta q\right) _{sea}<0.
\end{eqnarray*}
Furthermore, the recent SMC data on the semi-inclusive DIS scattering\cite
{smc-semi} tentatively suggested $\Delta _{\bar{u}}\simeq $ $\Delta _{\bar{d}%
}\simeq 0.\;$Thus while the inclusive experiments point to a negatively
polarized quark sea, the semi-inclusive result indicates that the antiquarks
in this sea are not polarized.

The NMC measurement of the muon scatterings off proton and neutron targets
shows that the Gottfried sum rule is violated\cite{nmc}. It has been
interpreted as showing $\overline{d}>\overline{u}$ in the proton. This
conclusion has been confirmed by the asymmetry measurement (by NA51\cite
{na51}) in the Drell-Yan processes with proton and neutron targets, which
yield, at a specific quark momentum fraction value ($x=0.18$), the result of 
$\overline{d}\simeq 2\overline{u}$ in Eq.(\ref{na51result}).

To summarize, the quark-sea is ''observed'' to be very different from nQS.
It has the following flavor and spin structures: 
\[
\begin{tabular}{cc}
& \underline{Observed features of the quark sea} \\ 
$flavor:$ & $\text{\ }\overline{d}>\overline{u}\;\;\;\text{ and \ \ \ }%
\overline{s}\neq 0$ \\ 
$spin:$ & $\left( \Delta q\right) _{sea}<0$\ \ \ yet\ \ \ $\Delta _{\bar{q}%
}\simeq 0.$%
\end{tabular}
\]
By the statement of $\overline{s}\neq 0,$ we mean that OZI rule is not
operative for the strange quark. Recall our discussion in Sec. 1, this means
that the \emph{couplings} for the $\left( s\bar{s}\right) $-pair production
or annihilation are not suppressed, although the process may well be
inhibited by phase space factors. Namely, a violation of the OZI rule
implies that, to the extent one can ignore the effects of SU(3) breaking,
there should be significant amount of $\left( s\bar{s}\right) $-pairs in the
proton.

\subsection{The chiral quark idea of Georgi and Manohar}

Let us start with theoretical attempts to understand the flavor asymmetry of
\ $\overline{d}>\overline{u}$ in the proton's quark sea$:$

\emph{Pauli exclusion principle} and the $u$-$d$ valence-quark asymmetry in
the proton would bring about a suppression of the gluonic production of $%
\bar{u}^{\prime }\,$s (versus $\bar{d}^{\prime }\,$s). Thus it has been
pointed out long ago\cite{FF-pauli} that \ $\overline{d}=\overline{u}$ would
not strictly hold even in perturbative QCD due to the fact the $u^{\prime }s$
and $d^{\prime }$s in the $q\bar{q}$ pairs must be antisymmetrized with the $%
u^{\prime }s$ and $d^{\prime }$s of the valence quarks. This mechanism is
difficult to implement as the parton picture is intrinsically incoherent. In
short, the observed large flavor-asymmetry reminds us once more that the
study of quark sea is intrinsically a non-perturbative problem.

\emph{Pion cloud mechanism}\cite{Henley} is another idea to account for the
observed $\overline{d}>\overline{u}$ asymmetry. The suggestion is that the
lepton probe also scatters off the pion cloud surrounding the target proton
(the Sullivan process\cite{Sullivan}), and the quark composition of the pion
cloud is thought to have more $\bar{d}\,$s than $\bar{u}\,$s. There is an
excess of $\pi ^{+}$ (hence $\bar{d}\,^{\prime }$s) compared to $\pi ^{-}$,
because $p\rightarrow n+\pi ^{+}$, but not a $\pi ^{-}$ if the final states
are restricted nucleons. (Of course, $\pi ^{0}$s has $\overline{d}=\overline{%
u}.$) However, it is difficult to see why the long distance feature of the
pion cloud surrounding the proton should have such a pronounced effect on
the DIS processes, which should probe the \emph{interior} of the proton, and
also this effect should be significantly reduced by the emissions such as $%
p\rightarrow \Delta ^{++}+\pi ^{-},\;etc.$

Nevertheless, we see that the pion cloud idea does offer the possibility to
getting a significant $\overline{d}>\overline{u}$ asymmetry. One can improve
upon this approach by adopting the chiral quark idea of Georgi and Manohar%
\cite{mgtheor} so that there is such a mechanism operating in the \emph{%
interior} of the hadron. Here a set of internal Goldstone bosons couple
directly to the constituent quarks \emph{inside} the proton. In the
following, we will first review the chiral quark model which was invented to
account for the successes of simple constituent quark model.

\paragraph{The chiral quark idea}

Although we still cannot solve the non-perturbative QCD, we are confident it
must have the features of (1) color confinement, and (2) spontaneous
breaking of chiral symmetry.

\emph{Confinement:\ }Asymptotic freedom $\alpha _{s}\left( Q\right) 
\stackunder{Q\rightarrow \infty }{\longrightarrow }0$ suggests that the
running coupling increases at low momentum-transfer and long distance, and $%
\alpha _{s}\left( \Lambda _{QCD}\right) \simeq 1$ is responsible for the
binding of quarks and gluons into hadrons. Experimental data indicates a
confinement scale at 
\begin{equation}
\Lambda _{QCD}\simeq 100\;\text{to }300\,MeV.  \label{qcd-scale}
\end{equation}

\emph{Chiral symmetry breaking: }There are three light quark flavors, $%
m_{u,d,s}<\Lambda _{QCD}.\;$In the approximation of $m_{u,d,s}=0,$ the QCD
Lagrangian is invariant under the independent $SU(3)$ transformations of the
left-handed and right-handed light-quark fields. Namely, the QCD Lagrangian
has a global symmetry of $SU(3)_{L}\times SU(3)_{R}$. If it is realized in
the normal Wigner mode, we should expect a chirally degenerate particle
spectrum: an octet of scalar mesons having approximately the same masses as
the octet pseudoscalar mesons, spin $\frac12$$^{-}\;$baryon octet degenerate
with the familiar $\frac12$$^{+}\;$baryon octet, etc. The absence of such
degeneracy suggests that the symmetry must be realized in the
Nambu-Goldstone mode: the QCD vacuum is not a chiral singlet and it
possesses a set of quark condensate $\left\langle 0\left| \bar{q}q\right|
0\right\rangle \neq 0.\;$Thus the symmetry is spontaneously broken 
\[
SU(3)_{L}\times SU(3)_{R}\rightarrow SU(3)_{L+R} 
\]
giving rise to an octet of approximately massless pseudoscalar mesons, which
have successfully been identified with the observed $\left( \pi ,\,K,\,\eta
\right) $ mesons.

The QCD Lagrangian is also invariant under the axial U(1) symmetry, which
would imply the ninth GB $m_{\eta ^{\prime }}\simeq m_{\eta }.\,$But the
existence of axial anomaly breaks the symmetry and in this way the eta prime
picks up an extra mass.

Both confinement and chiral symmetry breaking are non-perturbative QCD
effects. However, they have different physical origin; hence, it's likely
they have different distance scales. It is quite conceivable that as energy $%
Q$ decreases, but before reaching the confinement scale, $\alpha _{s}\left(
Q\right) $ has already increased to a sufficient size that it triggers
chiral symmetry breaking ($\chi SB$). This scenario 
\begin{equation}
\Lambda _{QCD}<\Lambda _{\chi SB}\simeq 1\,GeV.  \label{csb-scale}
\end{equation}
is what Georgi and Manohar have suggested to take place. The numerical value
is a guesstimate from the applications of chiral perturbation theory: $%
\Lambda _{\chi SB}\simeq 4\pi f_{\pi }$ with $f_{\pi }$ being the pion decay
constant. Because of this separation of the two scales, in the \emph{interior%
} of hadron, 
\[
\Lambda _{QCD}<Q<\Lambda _{\chi SB}, 
\]
the Goldstone boson$\ $(GB) excitations already become relevant (we call them%
\emph{\ internal GBs}), and the important effective DOFs are quarks, gluons
and internal GBs. In this energy range the quarks and GBs propagate in the
QCD vacuum which is filled with the $\bar{q}q$ condensate: the interaction
of a quark with the condensate will cause it to gain an extra mass of $%
\simeq 350\,MeV.$ This is the chiral quark model explanation of the large
constituent quark mass, (much in the same manner how all leptons and quark
gain their Lagrangian masses in the standard electroweak theory). The
precise relation between the internal and the physical GBs is yet to be
understood. The non-perturbative strong gluonic color interactions are
presumably responsible for all these effects. But once the physical
description is organized in terms of the resultant constituent quarks and
internal GBs (in some sense, the most singular parts of the original gluonic
color interaction) it is possible that the remanent interactions between the
gluons and quarks/GBs are not important. (The analogy is with quasiparticles
in singular potential problems in ordinary quantum mechanics.) Thus in our $%
\chi $QM description we shall ignore the gluonic degrees of freedom
completely.

\begin{description}
\item[\textbf{Remark }]  One may object to this omission of the gluonic DOF
on ground that the one gluon exchange\cite{DRGG} is needed to account for
the spin-dependent contributions to the hadronic mass as discussed in Sec.
1. However, in the $\chi $QM the constituent quarks interact through the
exchange of GBs. The axial couplings of the GB-quark couplings reduce to the
same $\frac{\mathbf{s}_{i}\cdot \mathbf{s}_{j}}{m_{i}m_{j}}$ effective terms
as the gluonic exchange couplings. For a more thorough discussion of hadron
spectroscopy in such a chiral quark description see recent work by Glozman
and Riska\cite{Glozman}.
\end{description}

\subsection{Flavor-spin structure of the nucleon}

In the chiral quark model the most important effective interactions in the
hadron interior for $Q<1\,GeV$ are the couplings of internal GBs to
constituent quarks. The phenomenological success of this model requires that
such interactions being feeble enough that perturbative description is
applicable. This is so, even though the underlying phenomena of spontaneous
chiral symmetry breaking and confinement are, obviously, non-perturbative.

\subsubsection{Chiral quark model with an octet of Goldstone bosons}

Bjorken\cite{bj-chiral}, Eichten, Hinchliffe and Quigg\cite{ehq} are the
first ones to point out that the observed flavor and spin structures of
nucleon are suggestive of the chiral quark features. In this model the
dominant process is the fluctuation of a valence quark $q$ into quark $%
q^{\prime }$ plus a Goldstone boson, which in turn is a $\left( q\bar{q}%
^{\prime }\right) $ system: 
\begin{equation}
q_{\pm }\longrightarrow GB+q_{\mp }^{\prime }\longrightarrow \left( q\bar{q}%
^{\prime }\right) _{0}q_{\mp }^{\prime }.  \label{cqm-basic}
\end{equation}
This basic interaction causes a modification of the spin content because a
quark changes its helicity (as indicated by the subscripts) by emitting a
spin-zero meson in P-wave. It causes a modification of the flavor content
because the GB fluctuation, unlike gluon emission, is flavor dependent.

In the absence of interactions, the proton is made up of two $u$ quarks and
one $d$ quark. We now calculate the proton's flavor content after any one of
these quarks turns into part of the quark sea by ``disintegrating'', \emph{%
via} GB emissions, into a quark plus a quark-antiquark pair.

Suppressing all the space-time structure and only displaying the flavor
content, the basic GB-quark interaction vertices are given by 
\begin{eqnarray}
\mathcal{L}_{I} &=&g_{8}\bar{q}\Phi q=g_{8}\left( 
\begin{array}{ccc}
\bar{u} & \bar{d} & \bar{s}
\end{array}
\right) \left( 
\begin{array}{ccc}
\frac{\pi ^{0}}{\sqrt{2}}+\frac{\eta }{\sqrt{6}} & \pi ^{+} & K^{+} \\ 
\pi ^{-} & -\frac{\pi ^{0}}{\sqrt{2}}+\frac{\eta }{\sqrt{6}} & K^{0} \\ 
K^{-} & \bar{K}^{0} & -\frac{2\eta }{\sqrt{6}}
\end{array}
\right) \left( 
\begin{array}{c}
u \\ 
d \\ 
s
\end{array}
\right)  \nonumber \\
&=&g_{8}\left[ \bar{d}\pi ^{-}+\bar{s}K^{-}+\bar{u}\left( \frac{\pi ^{0}}{%
\sqrt{2}}+\frac{\eta }{\sqrt{6}}\right) \right] u+...  \label{octet-coupling}
\end{eqnarray}
Thus after one emission of the \emph{u} quark wavefunction has the
components 
\begin{equation}
\Psi \left( u\right) \sim \left[ d\pi ^{+}+sK^{+}+u\left( \frac{\pi ^{0}}{%
\sqrt{2}}+\frac{\eta }{\sqrt{6}}\right) \right] ,  \label{uWF}
\end{equation}
which can be expressed entirely in terms of quark contents by using $\pi
^{+}=u\bar{d},$ and $K^{+}=u\bar{s},\;etc.\;$Since $\pi ^{0}$ and $\eta $
have the same quark contents, we can add their amplitudes coherently so that 
\begin{equation}
\left( \frac{\pi ^{0}}{\sqrt{2}}+\frac{\eta }{\sqrt{6}}\right) =\frac{2}{3}u%
\bar{u}-\frac{1}{3}d\bar{d}+\frac{1}{3}s\bar{s}  \label{add-pi-eta}
\end{equation}
Square the wavefunction we the obtain the probability of the transitions:
for example, 
\begin{equation}
\text{Prob}\left[ u_{+}\rightarrow \pi ^{+}d_{-}\rightarrow \left( u%
\overline{d}\right) _{0}d_{-}\right] \equiv a,  \label{prob-a}
\end{equation}
which will be used to set the scale for other emissions. At this stage we
shall assume SU(3) symmetry. Hence all processes have the same phase space,
and are proportional to the same probability $a\propto \left| g_{8}\right|
^{2}.$ The specific values are listed in the 3rd column of Table 2. The 2nd
column is the isospin counter-part obtained by the exchange of $%
u\leftrightarrow d:$%
\[
\begin{tabular}{|c|c|c|c|}
\hline\hline
$u_{+}\rightarrow $ & $d_{+}\rightarrow $ & SU(3) sym prob & broken U(3) prob
\\ 
&  & octet GB & nonet GB \\ \hline
$u_{+}\rightarrow \left( u\overline{d}\right) _{0}d_{-}$ & $d_{+}\rightarrow
\left( d\overline{u}\right) _{0}u_{-}$ & $a$ & $a$ \\ 
$u_{+}\rightarrow \left( u\overline{s}\right) _{0}s_{-}$ & $d_{+}\rightarrow
\left( d\overline{s}\right) _{0}s_{-}$ & $a$ & $\epsilon ^{2}a$ \\ 
$u_{+}\rightarrow \left( u\overline{u}\right) _{0}u_{-}$ & $d_{+}\rightarrow
\left( d\overline{d}\right) _{0}d_{-}$ & $\frac{4}{9}a$ & $\left( \frac{%
\delta +2\zeta +3}{6}\right) ^{2}a$ \\ 
$u_{+}\rightarrow \left( d\overline{d}\right) _{0}u_{-}$ & $d_{+}\rightarrow
\left( u\overline{u}\right) _{0}d_{-}$ & $\frac{1}{9}a$ & $\left( \frac{%
\delta +2\zeta -3}{6}\right) ^{2}a$ \\ 
$u_{+}\rightarrow \left( s\overline{s}\right) _{0}u_{-}$ & $d_{+}\rightarrow
\left( s\overline{s}\right) _{0}d_{-}$ & $\frac{1}{9}a$ & $\left( \frac{%
\zeta -\delta }{3}\right) ^{2}a$ \\ \hline\hline
\end{tabular}
\]

\begin{center}
\underline{Table 2}\ \ $\chi QM$ transition probabilities calculated in
models with an octet GB in the SU(3) symmetric limit and with nonet GB and
broken-U(3) breakings.
\end{center}

\paragraph{Flavor content calculation}

From Table 2, one can immediately read off the antiquark number $\bar{q}$ in
the proton after one emission of GB by the initial valence quarks $\left(
2u+d\right) $ in the proton: 
\begin{eqnarray}
\bar{u} &=&2\times \frac{4}{9}a+a+\frac{1}{9}a=2a  \label{octet-qbar} \\
\bar{d} &=&2\times \left( a+\frac{1}{9}a\right) +\frac{4}{9}a=\frac{8}{3}a 
\nonumber \\
\bar{s} &=&2\times \left( a+\frac{1}{9}a\right) +\left( a+\frac{1}{9}%
a\right) =\frac{10}{3}a  \nonumber
\end{eqnarray}
Since the quark and antiquark numbers must equal in the quark sea, we have
the quark numbers in the proton: 
\begin{equation}
u=2+\bar{u},\;\;\;\;\;d=1+\bar{d},\;\;\;\;\;s=\bar{s}.  \label{q-content}
\end{equation}

\paragraph{Spin content calculation}

GB emission will flip the helicity of the quark as indicated in the basic
process of (\ref{cqm-basic}), while the quark-antiquark pair produced
through the GB channel are unpolarized: 
\begin{equation}
\psi \left( GB\right) =\frac{1}{\sqrt{2}}\left[ \psi \left( q_{+}\right)
\psi \left( \bar{q}_{-}^{\prime }\right) -\psi \left( q_{-}\right) \psi
\left( \bar{q}_{+}^{\prime }\right) \right] .  \label{spin0comb}
\end{equation}
One of the first $\chi QM$ predictions about the spin structure is that, to
the leading order, the antiquarks are not polarized: 
\begin{equation}
\Delta _{\bar{q}}=\bar{q}_{+}-\bar{q}_{-}=0.  \label{antiq-pol}
\end{equation}
Before GB emissions as in (\ref{cqm-basic}), the proton wavefunction is
given by Eq.(\ref{spinWF}) giving the spin-dependent quark numbers in Eq.(%
\ref{spin-density}). Now from the 3rd column in Table 2, we can read off the
first-order probabilities: 
\begin{equation}
P_{1}\left( u_{+}\rightarrow d_{-}\right) =a\;\;\;P_{1}\left(
u_{+}\rightarrow s_{-}\right) =a\;\;\;P_{1}\left( u_{+}\rightarrow
u_{-}\right) =\frac{2}{3}a,  \label{prob1}
\end{equation}
or write this in a more compact notation as 
\begin{equation}
P_{1}\left( u_{+}\rightarrow \right) =(d_{-}+s_{-}+\frac{2}{3}u_{-})a.
\label{prob1compact}
\end{equation}
From this we can also immediately obtain the related probabilities of $%
P_{1}\left( u_{-}\rightarrow \right) ,$ $P_{1}\left( d_{+}\rightarrow
\right) ,$ and $P_{1}\left( d_{-}\rightarrow \right) .$ The sum of the three
terms in Eq.(\ref{prob1}) being $\frac{8}{3}a,$ the probability of \emph{no
GB emission} must then be $\left( 1-\frac{8}{3}a\right) $. Combining the 0th
and 1st order terms of Eqs.(\ref{spin-density}) and (\ref{prob1compact}), we
find the spin-dependent quark densities (coefficients in front of $q_{\pm }$%
): 
\begin{eqnarray*}
&&\left( 1-\frac{8}{3}a\right) \left( \frac{5}{3}u_{+}+\frac{1}{3}u_{-}+%
\frac{1}{3}d_{+}+\frac{2}{3}d_{-}\right) +\frac{5}{3}(d_{-}+s_{-}+\frac{2}{3}%
u_{-})a \\
&&+\frac{1}{3}(d_{+}+s_{+}+\frac{2}{3}u_{+})a+\frac{1}{3}(u_{-}+s_{-}+\frac{2%
}{3}d_{-})a+\frac{2}{3}(u_{+}+s_{+}+\frac{2}{3}d_{+})a
\end{eqnarray*}
Together with Eq.(\ref{antiq-pol}), we can then calculate the quark
polarization in the proton $\Delta q=\Delta _{q}+\Delta _{\bar{q}}=\Delta
_{q}=q_{+}-q_{-}:$%
\begin{equation}
\Delta u=\frac{4}{3}-\frac{37}{9}a,\;\;\;\;\;\;\;\Delta d=-\frac{1}{3}-\frac{%
2}{9}a,\;\;\;\;\;\;\;\Delta s=-a.  \label{octet-delq}
\end{equation}
In order to account for the NMC data of Eq.(\ref{ubar-dbar}) by $\bar{u}-%
\bar{d}=-\frac{2}{3}a$ as in Eq.(\ref{octet-qbar}), we need a probability of 
$\;a\simeq 0.22.\;$But such a large probability would lead to spin content
description that can at best be described as fair. For example it give a
negative-valued total quark value of $\Delta \Sigma =1-16a/3\simeq -0.17,\,$%
which is clearly incompatible with the current phenomenological values in
Eq.(\ref{delq-expt}) --- although it was still marginally consistent with
the original EMC value when this calculation was first performed\cite{ehq}.
Also, the antiquark numbers in Eq.(\ref{octet-qbar}) leads to a fixed ratio
of $\bar{u}/\bar{d}=0.75,$ which is to be compared to the NA51 result of $%
0.51,$ as given in Eq.(\ref{na51result}).

\subsubsection{Chiral quark model with a nonet of Goldstone bosons}

We have proposed\cite{CL95} a broken-U(3) version of the chiral quark model
with the inclusion of the ninth GB, the $\eta ^{\prime }$meson.

Besides the phenomenological considerations discussed above, we have also
been motivated to modify the original $\chi QM$ by the following theoretical
considerations. It is well-known that 1/N$_{color}$ expansion can provide us
with a useful guide to study non-perturbative QCD. In the leading 1/N$%
_{color}$ expansion (the planar diagrams), there are \emph{nine}\textbf{\ }%
GBs with an U(3) symmetry. Thus from this view point we should include the
ninth GB, the $\eta ^{\prime }$ meson. However we also know that if we stop
at this order, some essential physics would have been missed: At the planar
diagram level there is no axial anomaly and $\eta ^{\prime }$ would have
been a \emph{bona fide} GB. Also, it has been noted by Eichten \emph{et al.}%
\cite{ehq} that an unbroken U(3) symmetry would also lead to the
phenomenologically unsatisfactory feature of a flavor-symmetric sea: $\bar{u}%
=\bar{d}=\bar{s}$, which clearly violates the experimental results of Eqs.(%
\ref{ubar-dbar}) and (\ref{na51result}). Mathematically, this flavor
independence comes about as follows. Equating the coupling constants $%
g_{8}=g_{1}$ in the vertex which generalizes the coupling in Eq.(\ref
{octet-coupling}) 
\begin{equation}
\mathcal{L}_{I}=g_{8}\sum_{i=1}^{8}\bar{q}\lambda _{i}\phi _{i}q+\sqrt{\frac{%
2}{3}}g_{1}\bar{q}\eta ^{\prime }q  \label{nonet-coupling}
\end{equation}
($\lambda _{i}\phi _{i}=\Phi $ with $\lambda _{i\text{ }}$being the
Gell-Mann matrices) and squaring the amplitude, one obtains the probability
distribution of 
\begin{equation}
\sum_{i=1}^{8}\left( \bar{q}\lambda _{i}q\right) \left( \bar{q}\lambda
_{i}q\right) +\frac{2}{3}\left( \bar{q}q\right) \left( \bar{q}q\right)
\label{U3sum}
\end{equation}
which has the index structure as 
\begin{equation}
\sum_{i=1}^{8}\left( \lambda _{i}\right) _{ab}\left( \lambda _{i}\right)
_{cd}+\frac{2}{3}\delta _{ab}\delta _{cd}=2\delta _{ad}\delta _{bc}
\label{GMmatrix-id}
\end{equation}
where we have use a well-known identity of the Gell-Mann matrices to obtain
the equality. This clearly shows the flavor independence nature of the
result.

\paragraph{Calculation in the degenerate mass limit}

All this shows that we should include the ninth GB but, at the same time, it
is crucial that this resultant flavor-U(3) symmetry be broken. In our
earlier publication\cite{CL95} we have implemented this breaking in the
simplest possible manner by simply allowing the octet and singlet couplings
be different. Namely, in the first round calculation, we stayed with
approximation of $m_{n}=m_{s}$ and a degenerate octet GBs. In this way we
were able to show that with a choice of 
\begin{equation}
\zeta \equiv \frac{g_{1}}{g_{8}}\simeq -1  \label{neg-zeta}
\end{equation}
this broken U(3) $\chi QM$ can account for much of the observed spin and
flavor structure, see Column-5 in Table 3.

Our calculation has been performed in the SU(3) symmetric limit (\emph{i.e.}
assumed all phase space factors are the same). In this spirit we have chosen
to work with $\left| g_{1}\right| =\left| g_{8}\right| $ . The relative
negative sign is required primarily to yield an antiquark relation of $\bar{d%
}\simeq 2\bar{u}:$ as the model calculation gives a ratio 
\begin{equation}
\bar{u}/\bar{d}=\frac{\zeta ^{2}+2\zeta +6}{\zeta ^{2}+8}.  \label{sym-udbar}
\end{equation}
Therefore, the experimental value of Eq.(\ref{na51result}) implies a
negative coupling ratio : $-4.3<\zeta <-0.7.\;$We remark that the relative
sign of the couplings is physically relevant because of the interference
effects when we coherently add the $\eta ^{\prime }$ contribution to those
by $\eta $ and $\pi ^{0}.$ After fixing this ratio, there is only one
parameter $a$ that we can adjust to yield a good fit. It is gratifying that $%
a=0.11$ is indeed small, fulfilling our hope that once the singular features
of the nonperturbative phenomenon of spontaneous symmetry breaking are
collected in the GB degrees of freedom, the remanent dynamics among these
particles is perturbative in nature.

It should also be noted that we have compared these SU(3) symmetric results
to phenomenological values which have been extracted after using the SU(3)
symmetry relations as well. For example the result in Eq.(\ref{delq-expt})
have been extracted after using the SU(3) symmetric F/D ratio for hyperon
decays as in Eq.(\ref{del8}). Similarly, we obtained a strange quark
fraction value $F\left( s\right) \simeq 0.19$ very close to that given in
Eq.(\ref{sfrac}) which was deduced from $\sigma _{\pi N}$ and an SU(3)
symmetric F/D ratio for baryon masses, Eq.(\ref{M8}). Agreements are in the
20\% to 30\% range, indicating that the broken-U(3) chiral picture is,
perhaps, on the right track.

\paragraph{SU(3) and axial-U(1) breaking effects}

The quark mass difference $m_{s}>m_{n},$ and thus the GB nondegeneracy ,
would affect the phase space factors for various GB emission processes. Such
SU(3) breaking effects will be introduced\cite{song}\cite{CL97} in the
amplitudes for GB emissions, simply through the insertion of suppression
factors: $\epsilon $ for kaons, $\delta $ for eta, and $\zeta $ for eta
prime mesons, as these strange quark bearing GB's are more massive than the
pions. Thus the probability $a\propto $ $\left| g_{8}\right| ^{2}$are
modifies for processes involving strange quarks, as shown in the last column
of Table 2. The suppression factors enter into the probabilities for $%
u_{+}\rightarrow \left( u\overline{u}\right) _{0}u_{-}$ and $%
u_{+}\rightarrow \left( d\overline{d}\right) _{0}u_{-}$ processes, \emph{etc.%
} because they also receive contributions from the $\eta $ and $\eta
^{\prime }$ GBs. Following the same steps as those in Eqs.(\ref
{octet-coupling}) to (\ref{prob-a}), we obtain the probabilities as listed
in the 4th column of Table 2. In this way the following results are
calculated: 
\begin{eqnarray}
\overline{u} &=&\frac{1}{12}\left[ \left( 2\zeta +\delta +1\right)
^{2}+20\right] a,  \label{ubar} \\
\overline{d} &=&\frac{1}{12}\left[ \left( 2\zeta +\delta -1\right)
^{2}+32\right] a,  \label{dbar} \\
\overline{s} &=&\frac{1}{3}\left[ \left( \zeta -\delta \right)
^{2}+9\epsilon ^{2}\right] a.  \label{sbar}
\end{eqnarray}
and 
\begin{eqnarray}
\Delta u &=&\frac{4}{3}-\frac{21+4\delta ^{2}+8\zeta ^{2}+12\epsilon ^{2}}{9}%
a  \label{delu} \\
\Delta d &=&-\frac{1}{3}-\frac{6-\delta ^{2}-2\zeta ^{2}-3\epsilon ^{2}}{9}a
\label{deld} \\
\Delta s &=&-\epsilon ^{2}a  \label{dels}
\end{eqnarray}
In the limit of $\zeta =0$ (\emph{i.e. }no $\eta ^{\prime }$) and $\epsilon
=\delta =1$ (no suppression in the degenerate mass limit) these results are
reduced to those of Eqs.(\ref{octet-qbar}) and (\ref{octet-delq}).

Results of the numerical calculation are given in the last column in Table
3. Again our purpose is not so much as finding the precise best-fit values,
but using some simple choice of parameters to illustrate the structure of
chiral quark model. For more detail of the parameter choice, see Ref.\cite
{CL97}.

\begin{center}
\begin{tabular}{|c|c|c|c|c|c|}
\hline\hline
&  &  & Naive & $\chi $QM & $\chi $QM \\ 
& Phenomenological & Eq. & QM & SU$_{3}$ sym & brok'n SU$_{3}$ \\ 
& value & \# &  & $\epsilon =\delta =$ & $\epsilon =\delta =$ \\ 
&  &  &  & $-\zeta =1$ & $-2\zeta =0.6$ \\ 
&  &  & $a=0$ & $a=0.11$ & $a=0.15$ \\ \hline
&  &  &  &  &  \\ 
$\overline{u}-\overline{d}$ & $0.147\pm 0.026$ & (\ref{ubar-dbar}) & $0\,?$
& $0.146$ & $0.15$ \\ 
$\bar{u}/\bar{d}$ & $\left( 0.51\pm 0.09\right) _{x=0.18}$ & (\ref
{na51result}) & $1?$ & $0.56$ & $0.63$ \\ 
$2\bar{s}/\left( \bar{u}+\bar{d}\right) $ & $\simeq 0.5$ &  & $0?$ & $1.86$
& $0.60$ \\ 
&  &  &  &  &  \\ 
$\sigma _{\pi N}:F\left( s\right) $ & $0.18\pm 0.06(\downarrow ?)$ & (\ref
{sfrac}) & $0?$ & $0.19$ & $0.09$ \\ 
$F\left( 3\right) /F\left( 8\right) $ & $0.23\pm 0.05$ & (\ref{f3f8-sqm}) & $%
\frac{1}{3}$ & $\frac{1}{3}$ & $0.22$ \\ 
&  &  &  &  &  \\ 
$g_{A}$ & $1.257\pm 0.03$ &  & $\frac{5}{3}$ & $1.12$ & $1.25$ \\ 
$\left( F/D\right) _{axial}$ & $0.575\pm 0.016$ &  & $\frac{2}{3}$ & $\frac{2%
}{3}$ & $0.57$ \\ 
$\left( 3F-D\right) _{a}$ & $0.60\pm 0.07\;\left( \downarrow ?\right) $ & (%
\ref{del8}) & $1$ & $0.67$ & $0.59$ \\ 
&  &  &  &  &  \\ 
$\Delta u$ & $0.82\pm 0.06$ &  & $\frac{4}{3}$ & $0.78$ & $0.85$ \\ 
$\Delta d$ & $-0.44\pm 0.06$ &  & $-\frac{1}{3}$ & $-0.33$ & $-0.40$ \\ 
$\Delta s$ & $-0.11\pm 0.06\;\left( \downarrow ?\right) $ & (\ref{delq-expt})
& $0$ & $-0.11$ & $-0.07$ \\ 
&  &  &  &  &  \\ 
$\Delta \bar{u},\;\Delta \bar{d}$ & $-0.02\pm \left( .11\right) $ & (\ref
{smc-semi-result}) &  & $0$ & $0$ \\ \hline\hline
\end{tabular}
\end{center}

\underline{Table 3}\ Comparison of $\chi QM$ with phenomenological values.
The 3rd column gives the Eq. numbers where these values are discussed. From
there one can also look up the reference for the source of these values.
Possible downward revision of the results by SU(3) breaking effects, as
discussed in the text, are indicated by the symbol $\left( \downarrow
?\right) .$ Those values with a question mark $\left( ?\right) $ in the 4th
column are not strictly the sQM predictions, but are the common expectations
of, what has been termed in Sec. 3.1, the ``naive quark sea''.\bigskip 

Since a SU(3) symmetric calculation would not alter the relative strength of
quantities belonging to the same SU(3) multiplet, our symmetric calculation
cannot be expected to improve on the naive quark model, \emph{i.e.} SU(6),
results such as the axial vector coupling ratio $F/D=2/3,$ which differs
significantly from the generally quoted phenomenological value of $%
F/D=0.575\pm 0.016.$ To account for this difference we must include the
SU(3) breaking terms: 
\begin{eqnarray}
\frac{F}{D} &=&\frac{\Delta u-\Delta s}{\Delta u+\Delta s-2\Delta d} 
\nonumber \\
&=&\frac{2}{3}\cdot \frac{6-a\left( 2\delta ^{2}+4\zeta ^{2}+\frac{1}{2}%
\left( 3\epsilon ^{2}+21\right) \right) }{6-a\left( 2\delta ^{2}+4\zeta
^{2}+9\epsilon ^{2}+3\right) }.  \label{ftod}
\end{eqnarray}
Similarly discussion holds for the $F/D$ ratio for the octet baryon masses.
Here we choose to express this in terms of the quark flavor fractions as
defined by Eqs.(\ref{frac-def}) and (\ref{frac38}): 
\begin{eqnarray}
\frac{F\left( 3\right) }{F\left( 8\right) } &=&\frac{F\left( u\right)
-F\left( d\right) }{F\left( u\right) +F\left( d\right) -2F\left( s\right) }=%
\frac{1+2\left( \bar{u}-\bar{d}\right) }{3+2\left( \bar{u}+\bar{d}-2\bar{s}%
\right) }  \nonumber \\
&=&\frac{1}{3}\cdot \frac{3+2a\left[ 2\zeta +\delta -3\right] }{3+2a\left[
2\zeta \delta +\frac{1}{2}\left( 9-\delta ^{2}-12\epsilon ^{2}\right)
\right] }.  \label{fratio}
\end{eqnarray}
In the SU(3) symmetry limit of $\delta =\epsilon =1,$ we can easily check
that Eqs.(\ref{ftod}) and (\ref{fratio}) reduce to their naive quark model 
\emph{i.e. SU(6) }values, independent of $a$ and $\zeta .\;$Again it is
gratifying to see, as displayed in Table 3, that $\chi QM$ has just the
right structure so the SU(3) breaking modifications make the correction in
the right direction.

\subsection{Strange quark content of the nucleon}

We have already discussed the number $\bar{s}$ of strange quarks in the
nucleon quark sea and their polarization $\Delta s$. They are examples of
the proton matrix elements of operators bilinear in the strange quark fields 
$\left\langle p\left| \bar{s}\Gamma _{i}s\right| p\right\rangle ,$ or in
general we need to study the quark bilinear matrix elements of $\left\langle
p\left| \bar{q}\Gamma _{i}q\right| p\right\rangle :$

\subsubsection{The scalar channel}

This operator counts the number of quarks plus the number of antiquarks in
the proton. In particular the octet components of $\left\langle p\left| \bar{%
u}u-\bar{d}d\right| p\right\rangle $ and $\left\langle p\left| \bar{u}u+\bar{%
d}d-2\bar{s}s\right| p\right\rangle $ can be gotten by SU(3) baryon mass
relations as we have shown in Eq.(\ref{frac38}). But in order to separate
out the individual terms, say $\left\langle p\left| \bar{s}s\right|
p\right\rangle $, we would need the singlet combination $\left\langle
p\left| \bar{u}u+\bar{d}d+\bar{s}s\right| p\right\rangle .\;$This is
provided by $\sigma _{\pi N}$ which is a linear combination of the singlet
and octet pieces. That is why a measurement of $\sigma _{\pi N}$ allows us
to do an SU(3) symmetric calculation of the strange quark content of the
nucleon.

We have emphasized that OZI violation means that the\ \emph{couplings} for $s%
\bar{s}$ pair creation and annihilation may not be suppressed even though
the phase space surely does not favor such processes. But the phase space
suppression is a ``trivial'' SU(3) breaking effect. Our chiral quark model
calculation is a concrete realization of this possibility: Had we ignored
the phase space difference, the GB-quark couplings are such that there would
be \emph{more }strange quark pairs than either of the nonstrange pairs in
the quark sea, as $s\bar{s}$ production by either $u$ or $d$ valence quarks
are not disfavored. Thus Eq.(\ref{octet-qbar}) give a relative quark
abundance in the quark sea of 
\begin{equation}
\bar{u}:\bar{d}:\bar{s}=3:4:5
\end{equation}
In the physical quark sea we do not really expect strange quark pairs to
dominate because of their production is suppressed by SU(3) breaking effects.

The $\chi $QM naturally suggests that the nucleon strange quark content $%
\bar{s}$ and polarization $\Delta s$ magnitude are lowered by the SU(3)
breaking effects as they are directly proportional to the amplitude
suppression factors, see Eqs.(\ref{sbar}) and (\ref{dels}). This is just the
trend found in the extracted phenomenological values. Gasser\cite{Gass81},
for instance, using a chiral loop model to calculate the SU(3) breaking
correction to the Gell-Mann-Okubo baryon mass formula, finds that the
no-strange-quark limit-value of $\left( \sigma _{\pi N}\right) _{0}$ is
modified from $25$ to $35\,MeV,$ [\emph{i.e.}$\,$the baryon mass $M_{8}$ in
Eq.(\ref{M8}) changed from $-200$ by SU(3) breakings to $-280\,MeV$], thus
the fraction $F\left( s\right) $ from $0.18$ to $0.10$. It matches closely
our numerical calculation with the illustrative parameters, see Table 3.

The strange quark content can also be expressed as the relative abundance of
the strange to non-strange quarks in the sea, which in this model is given
as 
\begin{equation}
\lambda _{s}\equiv \frac{\bar{s}}{\frac{1}{2}\left( \bar{u}+\bar{d}\right) }%
=4\frac{\left( \zeta -\delta \right) ^{2}+9\epsilon ^{2}}{\left( 2\zeta
+\delta \right) ^{2}+27}\simeq 1.6\epsilon ^{2}=0.6.  \label{lambda-s}
\end{equation}
This can be compared to the strange quark content as measured by the CCFR
Collaboration in their neutrino charm production experiment\cite{sbarexpt} 
\begin{equation}
\kappa \equiv \frac{\left\langle x\bar{s}\right\rangle }{\frac{1}{2}\left(
\left\langle x\bar{u}\right\rangle +\left\langle x\bar{d}\right\rangle
\right) }=0.477\pm 0.063,\;\;\;\text{where\ \ \ }\left\langle x\bar{q}%
\right\rangle =\int_{0}^{1}x\bar{q}\left( x\right) dx,  \label{kappa-s}
\end{equation}
which is often used in the global QCD reconstruction of parton distributions%
\cite{MRS}. The same experiment found no significant difference in the
shapes of the strange and non-strange quark distributions\cite{sbarexpt}: 
\[
\left[ x\bar{s}\left( x\right) \right] \;\propto \;\left( 1-x\right)
^{\alpha }\left[ \frac{x\bar{u}\left( x\right) +x\bar{d}\left( x\right) }{2}%
\right] ,
\]
with the shape parameter being consistent with zero, $\alpha =-0.02\pm 0.08.$
Thus, it is reasonable to use the CCFR findings to yield 
\begin{equation}
\lambda _{s}\simeq \kappa \simeq \frac{1}{2},  \label{s-content}
\end{equation}
which is a bit less than, but still compatible with, the value in Eq.(\ref
{lambda-s}).

Thus it is seen that the $\chi QM$ can yield a consistent account of the
strange quark content $\bar{s}$ of the proton sea. SU(3) breaking is the key
in reconciling the $\bar{s}$ value as measured in the neutrino charm
production and that as deduced from the pion nucleon sigma term$.$

\subsubsection{The axial-vector channel}

This operator measures the quark contribution to the proton spin. In
particular the octet components of 
\[
\left\langle p,s\left| \bar{u}\gamma _{\mu }\gamma _{5}u-\bar{d}\gamma _{\mu
}\gamma _{5}d\right| p,s\right\rangle 
\]
and 
\[
\left\langle p,s\left| \bar{u}\gamma _{\mu }\gamma _{5}u+\bar{d}\gamma _{\mu
}\gamma _{5}d-2\bar{s}\gamma _{\mu }\gamma _{5}s\right| p,s\right\rangle 
\]
can be gotten by SU(3) relations among the axial vector couplings of octet
baryon weak decays, Eqs.(\ref{del3}) and (\ref{del8}). But in order to
separate out the individual terms, say $\left\langle p,s\left| \bar{s}\gamma
_{\mu }\gamma _{5}s\right| p,s\right\rangle =2s_{\mu }\Delta s$, we would
need the singlet combination $\Delta u+\Delta d+\Delta s.\;$This is provided
by the first-moment of the structure function $\int g_{1}dx$ which is a
linear combination of the singlet and octet pieces. That is why a
measurement of $g_{1}\left( x\right) $ allows us to do an SU(3) symmetric
calculation of the strange quark content of the nucleon.

A number of authors have pointed out that phenomenologically extracted value
of strange quark polarization $\Delta s$ is sensitive to possible SU(3)
breaking corrections. While the effect is model-dependent, various
investigations\cite{su3br} -\cite{ucsd} all conclude that SU(3) breaking
correction tends to lower the magnitude of $\Delta s$. Some even suggested
the possibility of $\Delta s\simeq 0$ being consistent with experimental
data. Our calculation indicates that, while $\Delta s$ may be smaller than $%
0.10$, it is not likely to be significantly smaller than $0.05.$ To verify
this prediction, it is then important to pursue other phenomenological
methods that allow the extraction of $\Delta s$ without the need of SU(3)
relations.

Besides polarized DIS of charged lepton off nucleon, we can also use other
processes to determine $\Delta s.\;$In elastic neutrino-proton scattering,
we can separate out the axial form factors at zero momentum transfer, 
\begin{equation}
\left\langle p^{\prime }\left| \bar{q}\gamma _{\mu }\gamma _{5}q\right|
p\right\rangle =2\bar{u}\left( p^{\prime }\right) \left[ G_{1}^{\left(
q\right) }\left( Q^{2}\right) \gamma _{\mu }\gamma _{5}+\frac{q_{\mu }\gamma
_{5}}{2M_{p}}G_{2}^{\left( q\right) }\left( Q^{2}\right) \right] u\left(
p\right) .
\end{equation}
Thus we have $G_{1}^{\left( q\right) }\left( 0\right) =\Delta q.\;$The axial
vector matrix element arises from $Z$-boson exchange is proportion to 
\begin{equation}
\left\langle p^{\prime }\left| \bar{q}\mathcal{T}_{3}\gamma _{\mu }\gamma
_{5}q\right| p\right\rangle =\frac{1}{2}\left\langle p^{\prime }\left| \bar{u%
}\gamma _{\mu }\gamma _{5}u-\bar{d}\gamma _{\mu }\gamma _{5}d-\bar{s}\gamma
_{\mu }\gamma _{5}s\right| p\right\rangle
\end{equation}
where $\mathcal{T}_{3}$ is the 3rd component of the weak isospin operator.
The $\left\langle p\left| \bar{s}\gamma _{\mu }\gamma _{5}s\right|
p\right\rangle $ can be separated out because the first two terms are fixed
by the neutron axial coupling $g_{A}.$ Present data still have large error,
however they are consistent with a $\Delta s\neq 0$\cite{KapMano}$.$

The measurements of longitudinal polarization of $\Lambda $ in the
semi-inclusive process of $\bar{\nu}N\rightarrow \mu \Lambda +X$ \cite{WA58}
have also given support to a nonvanishing and negative $\Delta s.\,$ In this
connection, it's also important to pursue experimental measurements to check
the $\chi $QM prediction for a vanishing longitudinal polarization of $\bar{%
\Lambda}$ in the semi-inclusive processes reflecting the proton spin
property of $\Delta \bar{s}=0.$

\subsubsection{The pseudoscalar channel}

The nucleon matrix elements of the pseudoscalar quark density may be
physically relevant in Higgs coupling to the nucleon\cite{Cheng88}, \emph{%
etc. }Such operators may be related to the axial vector current operator
through the (anomalous) divergence equation (\ref{ano-div-q})\cite{CL89}. If
we define 
\begin{eqnarray}
\left\langle p\left| \bar{q}i\gamma _{5}q\right| p\right\rangle &=&\nu _{q}%
\bar{u}\left( p\right) i\gamma _{5}u\left( p\right)  \label{nu-q} \\
\left\langle p\left| trG^{\mu \nu }\widetilde{G}_{\mu \nu }\right|
p\right\rangle &=&-\Delta g2M_{p}\bar{u}\left( p\right) i\gamma _{5}u\left(
p\right)  \label{del-g}
\end{eqnarray}
so that the non-strange divergence equations may be written as 
\begin{eqnarray}
2M_{p}\Delta u &=&2m_{u}\nu _{u}-2M_{p}\left( \frac{\alpha _{s}}{2\pi }%
\Delta g\right)  \label{div-eq-nu} \\
2M_{p}\Delta d &=&2m_{d}\nu _{d}-2M_{p}\left( \frac{\alpha _{s}}{2\pi }%
\Delta g\right)  \nonumber
\end{eqnarray}
We would need one more condition in order to separate out the individual $%
m_{q}\nu _{q}$ terms. This may be obtained by saturation of the nonsinglet
channel by Goldstone poles. Let us recall that the Goldberger-Treiman
relation can be derived in the charge channel by the $\pi ^{\pm }$
pole-dominance of the pseudoscalar density. After taking the nucleon matrix
element of 
\[
\partial ^{\mu }\left( \bar{u}\gamma _{\mu }\gamma _{5}d\right) =\left(
m_{u}+m_{d}\right) \left( \bar{u}i\gamma _{5}d\right) 
\]
one obtains 
\begin{equation}
2M_{p}g_{A}=2f_{\pi }g_{\pi NN}+\mu _{\pm }  \label{gA-charged}
\end{equation}
where $\mu _{\pm }$ denotes the correction to the $\pi ^{\pm }$
pole-dominance, and is the correction to the $g_{A}$ as given by the GT
relation. Repeating the same for the neutral isovector channel 
\[
\partial ^{\mu }\left( \bar{u}\gamma _{\mu }\gamma _{5}u-\bar{d}\gamma _{\mu
}\gamma _{5}d\right) =2m_{u}\left( \bar{u}i\gamma _{5}u\right) -2m_{d}\left( 
\bar{d}i\gamma _{5}d\right) 
\]
we have 
\begin{equation}
2M_{p}g_{A}=2f_{\pi }g_{\pi NN}+\mu _{0}+\left( m_{u}-m_{d}\right) \left(
\nu _{u}+\nu _{d}\right)  \label{gA-neutral}
\end{equation}
Comparing these two expressions for $g_{A\text{ }}$ one concludes that the
singlet density $\left( \nu _{u}+\nu _{d}\right) $ must be small, on the
order of \emph{correction} to the GT expression of $g_{A}$. Assume that $\mu
_{0}\simeq \mu _{\pm },$ thus $\nu _{u}=-\nu _{d},$ we can solve the two
equations in (\ref{div-eq-nu}) in terms of the measured $\Delta u$ and $%
\Delta d$ given in Eq.(\ref{delq-expt})$.$%
\begin{equation}
m_{u}\nu _{u}=423\,MeV\;\;\;\;\;m_{d}\nu _{d}=-761\,MeV\;\;\;\;\;\frac{%
\alpha _{s}}{2\pi }\Delta g=-0.37.
\end{equation}
With these values we can also obtain 
\begin{equation}
m_{s}\nu _{s}=-451\,MeV\;.
\end{equation}
Because of the large strange quark masses $m_{s}$, this translates into
fairly small strange pseudoscalar matrix element of 
\[
\left\langle p\left| \bar{s}i\gamma _{5}s\right| p\right\rangle \simeq
0.03\left\langle p\left| \bar{d}i\gamma _{5}d\right| p\right\rangle . 
\]

\subsubsection{The vector channel}

Of course the vector charges 
\begin{equation}
Q^{i}=\int d^{3}xV_{0}^{i}\left( x\right) \;\;\;\;\;\;\;\;\text{with \ \ }%
V_{\mu }^{i}=\bar{q}\gamma _{\mu }\frac{\lambda ^{i}}{2}q
\end{equation}
are simply the generators of the flavor SU(3). In terms of the form factors
defined as 
\begin{equation}
\left\langle p^{\prime }\left| \bar{q}\gamma _{\mu }\frac{\lambda ^{i}}{2}%
q\right| p\right\rangle =\bar{u}\left( p\right) \left[ \gamma _{\mu
}F_{1}^{\left( q\right) }\left( Q^{2}\right) +i\frac{\sigma _{\mu \nu
}\left( p^{\prime }-p\right) ^{\nu }}{2M_{p}}F_{2}^{\left( q\right) }\left(
Q^{2}\right) \right] u\left( p\right)
\end{equation}
where $Q^{2}=\left( p^{\prime }-p\right) ^{2}$ is the momentum transfer, we
note that $\left. \left\langle p\left| \bar{q}\gamma _{\mu }\frac{\lambda
^{i}}{2}q\right| p\right\rangle \right| _{Q^{2}=0}$ are constrained by the
quantum numbers of the proton: 
\begin{equation}
F_{1}^{\left( u\right) }\left( 0\right) -F_{1}^{\left( d\right) }\left(
0\right) =1,\;\;\;\;\;\;F_{1}^{\left( s\right) }\left( 0\right) =0.
\end{equation}
However, the magnetic moment form factor $\;F_{2}^{\left( s\right) }\left(
0\right) $ needs not vanish. It is therefore interesting to measure this
quantity. This can be done through the observation of parity violation in
the scattering of charged-leptons off nucleon. The interference of the
photon-exchange and Z-boson-exchange diagrams can be used to isolate $%
F_{2}^{\left( s\right) }\left( 0\right) $. For detailed discussion, see Refs.%
\cite{McK-Beck}\cite{bates97}.

\subsection{Discussion}

In these lectures we have described an attempt to understand the nucleon
spin-flavor structure in the framework of a broken-U(3) chiral quark model.
The broad agreement obtained with simple schematic calculations, as
displayed in Table 3, has been quite encouraging. If this approach turns out
to be right, it just means that the familiar non-relativistic constituent
quark model is basically correct --- it only needs to be supplemented by a
quark sea generated by the valence quarks through their internal GB
emissions.

Because the couplings between GB and constituent quarks are not strong, we
can again use perturbation theory based on these non-perturbative degrees of
freedom --- even though the phenomena we are describing are non-perturbative
in terms of QCD Lagrangian quarks and gluons. Features such as $\bar{d}%
\;\simeq 2\bar{u}$ are seen to be clear examples of nonperturbative QCD
physics, as they are quite inexplicable in terms of a quark sea generated by
perturbative gluon emissions. (If one gets beyond the perturbative gluonic
picture, this $\bar{d}\;\neq \bar{u}$ property is not peculiar at all, as
the nucleon is not an isospin singlet and there is no reason to expect that
its quark sea should be an isospin singlet.)

In the case of the proton spin structure, because the most often discussed
theoretical interpretation is the possibility of a hidden gluonic
contribution, it has led some to think that other approaches, such as $\chi
QM,$ must be irrelevant. But the alternative theories are attempting a
different description by using different degrees of freedom. To be sure, the
QCD quarks and gluons are the most fundamental DOF. But we cannot insist on
using them for such non-perturbative problems as the hadron structure. An
analogy with the nucleon mass problem will illustrate our point.

The canonical approach to study the various quarks/gluon contributions to
the nucleon mass is through the energy-momentum trace anomaly equation\cite
{trace-ano}: 
\begin{equation}
\Theta _{\mu }^{\mu }=m_{u}\bar{u}u+m_{d}\bar{d}d+m_{s}\bar{s}s-\left( 11-%
\frac{2}{3}n_{f}\right) \frac{\alpha _{s}}{8\pi }trG^{\mu \nu }G_{\mu \nu }.
\end{equation}
Just like the more familiar axial vector anomaly equation, the naive
divergence is given by quark masses while the anomaly term is given by the
gluon field tensor. (Of course, here we are using the Lagrangian quark and
gluon fields.) When taken between the proton states, this equations yields 
\begin{equation}
M_{p}=m_{n}\left\langle p\left| \bar{u}u+\bar{d}d\right| p\right\rangle
+m_{s}\left\langle p\left| \bar{s}s\right| p\right\rangle +gluon\;term
\end{equation}
The first term is just the $\sigma _{\pi N}\simeq 45\,MeV$ representing a
tiny contribution by the nonstrange quarks, while the second term can also
be estimated\cite{Cheng88}: 
\begin{equation}
m_{s}\left\langle p\left| \bar{s}s\right| p\right\rangle =\frac{m_{s}\sigma
_{\pi N}}{2m_{n}}\left[ 1-\frac{3m_{n}}{m_{n}-m_{s}}\frac{M_{8}}{\sigma
_{\pi N}}\right] \simeq 250\,MeV.
\end{equation}
[Because this is an SU(3) calculation, the strange quark term is somewhat
overestimated.] One way or other, we see that most of the proton mass came
from the gluon term\cite{SVZ}. Heavy quark terms can also be included but
their contributions as explicit quark terms just cancel the corresponding
heavy quark loops in the gluon terms. In this sense they decouple\cite
{CL-rice}.

This led to an important insight: nucleon mass is mostly gluonic. But in
terms of the QCD quarks and gluons, it is difficult to say anything more.
That is why the description provided by the constituent quark model is so
important. In this picture much more details can be constructed: hyperfine
splitting, magnetic moments, etc.

The important point is that these two approaches are not mutually exclusive.
While the constituent quark model does not refer explicitly to gluon, the
above discussion suggests that it is the non-perturbative gluonic
interaction that brings about the large constituent quark masses. (In the $%
\chi QM$ this takes the form of quark interaction with the chiral condensate
of the QCD vacuum.) We believe that this complimentarity of the QCD and sQM
descriptions holds for the flavor-spin structure problem as well. The
non-perturbative features can be described much more succinctly if we use
the non-perturbative DOF of constituent quarks and internal GBs. Thus it is
quite possible that the statement of a significant gluonic contribution to
the proton spin and a correct description of spin structure by the $\chi QM$
can both be valid --- just the same physics expressed in two different
languages.\bigskip 

\begin{center}
\textbf{Acknowledgment}
\end{center}

L.F.L. would like to thank the organizers, in particular C.B. Lang, of the
Schladming Winter School for warm hospitality. His work is supported at CMU
by the U.S. Department of Energy (Grant No. DOE-ER/40682-127).

\end{document}